%% file: icml2026.tex
\documentclass{article}

\usepackage{microtype}
\usepackage{graphicx}
\usepackage{subcaption}
\usepackage{booktabs} 

\usepackage{hyperref}
\usepackage{caption} 

\newcommand{\benchmark}{\textsc{BugSourceBench}}
\newcommand{\ourslong}{\textsc{Anchored Self-Play (ASP)}}
\newcommand{\ours}{\textsc{ASP}}

\usepackage[accepted]{icml2026}

\usepackage{amsmath}
\usepackage{amssymb}
\usepackage{mathtools}
\usepackage{amsthm}

\usepackage[dvipsnames]{xcolor}
\usepackage[normalem]{ulem}
\usepackage{listings}
\usepackage[most]{tcolorbox}
\tcbuselibrary{listings,breakable,skins}
\usepackage{dblfloatfix} 
\usepackage{placeins} 

\lstdefinestyle{promptstyle}{
  basicstyle=\ttfamily\footnotesize,
  columns=fullflexible,
  breaklines=true,
  breakatwhitespace=false,
  keepspaces=true,
  showstringspaces=false,
  tabsize=2,
  numbers=left,
  numberstyle=\ttfamily\tiny\color{gray!70},
  numbersep=8pt,
  xleftmargin=10pt,   
  frame=none,
}

\newtcblisting{promptbox}[2][]{%
  breakable,
  enhanced,
  colback=gray!2,
  colframe=gray!35,
  boxrule=0.6pt,
  arc=2mm,
  left=8pt,right=8pt,top=6pt,bottom=6pt,
  listing only,
  listing options={style=promptstyle},
  title=\textbf{#2},
  fonttitle=\normalsize,
  coltitle=black,
  colbacktitle=gray!10,
  boxed title style={
    boxrule=0pt,
    arc=2mm,
    outer arc=2mm,
    left=6pt,right=6pt,top=3pt,bottom=3pt,
  },
  attach boxed title to top left={xshift=6pt,yshift=-2mm},
  borderline west={1.2pt}{0pt}{gray!45},
  drop shadow={black!10},
  #1
}

\definecolor{bugred}{RGB}{180,0,0}

\lstdefinestyle{py}{
  language=Python,
  basicstyle=\ttfamily\fontsize{6}{6}\selectfont,
  columns=fullflexible,
  breaklines=true,
  breakatwhitespace=true,
  showstringspaces=false,
  tabsize=1,
  aboveskip=0pt,
  belowskip=0pt,
  moredelim=**[is][\bfseries\color{bugred}]{@@}{@@}
}

\newtcolorbox{codebox}[2][]{%
  enhanced,
  breakable,
  colback=black!1,
  colframe=black!30,
  boxrule=0.3pt,
  arc=2pt,
  left=2pt,right=2pt,top=2pt,bottom=2pt,
  title=\textbf{#2},
  fonttitle=\footnotesize,
  coltitle=black,
  #1
}

\usepackage[capitalize,noabbrev]{cleveref}

\theoremstyle{plain}

\theoremstyle{definition}

\theoremstyle{remark}

\usepackage[disable,textsize=tiny]{todonotes}
\usepackage{enumitem}

\icmltitlerunning{Anchored Self-Play for Code Repair}

\begin{document}

\twocolumn[
  \icmltitle{Anchored Self-Play for Code Repair}

  \icmlsetsymbol{equal}{*}

  \begin{icmlauthorlist}
    \icmlauthor{Caroline Choi}{stanford}
    \icmlauthor{Zeyneb Kaya}{stanford}
    \icmlauthor{Shirley Wu}{stanford}
    \icmlauthor{Tengyu Ma}{stanford}
    \icmlauthor{Tatsunori Hashimoto}{stanford}
    \icmlauthor{Ludwig Schmidt}{stanford}
  \end{icmlauthorlist}

  \icmlaffiliation{stanford}{Department of Computer Science, Stanford University, Stanford, CA, USA}

  \icmlcorrespondingauthor{Caroline Choi}{cchoi1@stanford.edu}

  \icmlkeywords{code repair, program repair, self-play, large language models}

  \vskip 0.3in
]

\printAffiliationsAndNotice{}

\begin{abstract}
Code repair is an important capability for language models (LMs): given a buggy program and unit tests, an LM must produce a fixed program that passes the tests.
Because code repair data is limited, we aim to scale supervision by using an LM to generate bug--fix tasks.
We propose \emph{generator--fixer self-play}, in which a single model is trained with reinforcement learning to generate bugs and fix them.
As the fixer improves, the generator adapts to produce more difficult bugs, yielding an automatic curriculum.
To test whether this curriculum generalizes, we introduce \benchmark{}, a repair benchmark spanning realistic bug sources: bugs in human-written code, LM-generated code, and human-edited LM-generated code.
On \benchmark{}, we find that self-play drifts toward difficult but unrealistic bugs, improving on synthetic bugs but degrading on human-authored ones.
We propose \ourslong{}, which anchors self-play with a small reference set by adding a code-embedding similarity reward for generation and mixing reference bugs into fixer training.
Across bug sources, \ours{} achieves the best fix rates, improving average fix rate over standard self-play by $+24\%$ relative / $+7.0$ pp absolute, with gains on bugs from both LMs and humans.
\end{abstract}

\begin{figure*}[t]
    \centering
    \includegraphics[width=0.92\textwidth]{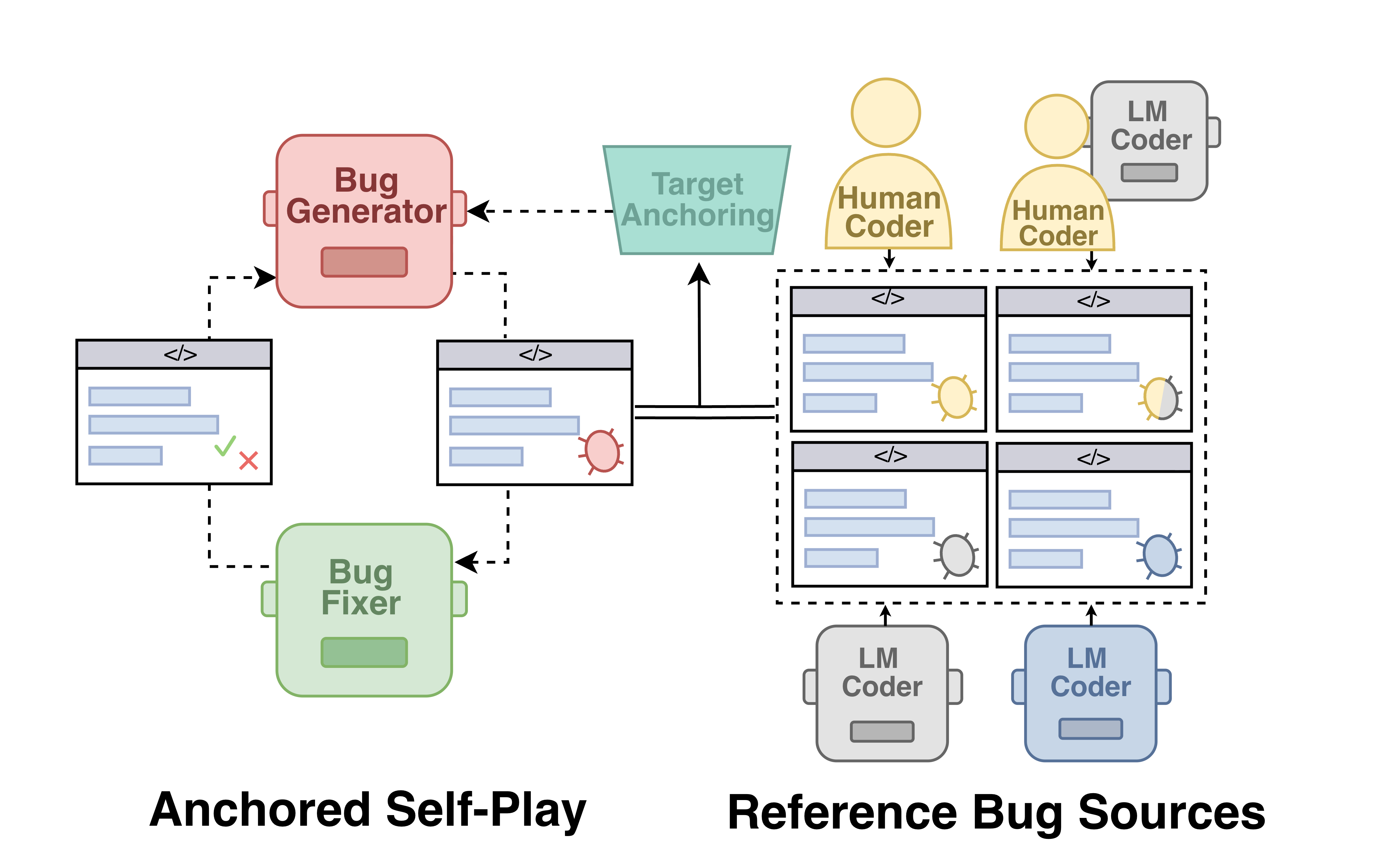}
    \caption{
    \textbf{Anchored self-play for code repair.}
    \emph{Left:} In generator--fixer self-play, the generator edits a correct program to produce a bug and the fixer repairs it; unit tests reward bug validity and repair correctness.
    Because unit tests verify pass/fail behavior but not realism, self-play can drift toward unrealistic test-failing bugs.
    \emph{Right:} \benchmark{} evaluates repair on the same programming tasks while varying the bug source, covering human-written bugs, human-edited LM bugs, and LM-generated bugs.
    \ourslong{} mitigates this drift by anchoring training to a small reference set, using an embedding-similarity reward for bug generation and reference-bug mixing for fixer training.
    }
    \label{fig:1}
\end{figure*}

\input{sections/2.introduction}

\input{sections/2.problem_formulation}
\input{sections/3.benchmark}

\input{sections/4.method}
\input{sections/5.experiments}

\input{sections/6.related_work}
\input{sections/7.conclusion}

\section*{Software and Data}
Code is available at
\url{https://github.com/cchoi1/anchored_self_play}.

\section*{Acknowledgements}
The authors thank Yangjun Ruan, Neil Band, Kaiyue Wen, Luke Bailey, Thomas Chen, and Arvind Mahankali for valuable discussions and feedback on this work. We also thank the anonymous reviewers for their constructive feedback. This work made use of computational resources provided by the Stanford Marlowe team~\citep{kapfer2025marlowe}. CC was supported by the National Science Foundation Graduate Research Fellowship Program under Grant No. DGE-2146755.

\input{sections/8.appendix}


\clearpage

\bibliography{references}
\bibliographystyle{icml2026}

\end{document}

%% file: sections/2.introduction.tex
\section{Introduction}

Code repair is an important capability for language models (LMs) used in programming workflows \citep{xu2022systematic,jimenez2023swe}.
Given a buggy program and accompanying unit tests, an LM must produce a fix that passes the tests.
However, high-quality, real-world repair data is limited \citep{just2014defects4j,widyasari2020bugsinpy,leGoues2015manybugs,madeiral2019bears,oliva2025spice}.

We ask whether high-quality supervision for code repair can be scaled using LMs to generate bug--fix tasks for training, with unit tests providing automatic verification.
We study an open-ended generation setting in which an LM can apply arbitrary text edits to correct code.
This allows for more diverse and adaptive synthetic data.
Ideally, as the fixer improves over training, the generator produces bugs that are increasingly challenging and realistic, forming an automatic curriculum.

We instantiate this idea with \emph{generator--fixer self-play} (Figure~\ref{fig:1}).
A single model is trained with reinforcement learning to alternate between generating a bug and fixing it.
The generator is rewarded for producing valid, appropriately difficult bugs (tests fail), and the fixer is rewarded for producing correct repairs (tests pass).

A key challenge in generator--fixer self-play is distribution drift.
Unit tests verify functional behavior, but they do not verify realism: many edits can break tests, but few resemble bugs encountered in practice.
As training progresses, the generator can drift toward difficult but unrealistic bugs, improving repair on self-generated tasks while degrading on real-world ones.
Prior work mitigates this by constraining bug generation to repository histories, deletions, or mutation operators \citep{wei2025ssr,forrest2009genprog,allamanis2021buglab}.
However, this limits the diversity and scale of synthetic training data and does not address the open-ended generation setting we consider here.

To evaluate whether self-play improves code repair beyond its own synthetic distribution, we introduce \benchmark{}, a  benchmark for cross-source repair generalization.
In LM-assisted programming, repair models may encounter bugs from multiple sources: human-written code, LM-generated code, and human edits to LM-generated code.
\benchmark{} therefore spans human-authored bugs, human edits of LM-generated bugs, and LM-generated bugs from weaker and stronger code models.
Crucially, \benchmark{} holds the repair task fixed while varying only the bug source.

On \benchmark{}, we find that standard self-play exhibits distribution drift: it improves on LM-generated bugs but regresses on human-authored bugs.
We propose \ourslong{} (\cref{fig:1}), which anchors self-play to a small reference set sampled from the target bug sources.
\ours{} uses this reference set in two complementary ways: a code-embedding similarity reward guides the generator toward target-like bugs, and reference-bug mixing exposes the fixer to realistic bugs throughout training.
Anchoring stabilizes self-play and improves cross-source repair: \ours{} achieves the best overall fix rate, improving average fix rate by $24\%$ relative / $+7.0$ pp absolute over standard self-play, with gains on both LM-generated bugs ($100\%$ relative / $+11$ pp absolute) and human-originated bugs ($7.1\%$ relative / $+3.4$ pp absolute).

Our contributions are:
\begin{itemize}[leftmargin=*, nosep, topsep=0pt, partopsep=0pt, itemsep=0.2pt, parsep=0pt]
  \item We formulate generator--fixer self-play for code repair and identify distribution drift as a key failure mode of unit-test-only self-play.
  \item We introduce and release \benchmark{}, a controlled multi-source benchmark spanning human-written, human-edited LM, and LM-generated bugs.
  \item We propose \ourslong{}, which reduces drift by combining embedding-similarity guidance for bug generation with reference-mixed fixer training.
\end{itemize}

%% file: sections/2.problem_formulation.tex
\section{Problem Formulation}
\label{sec:problem_formulation}

We study code repair with unit-test feedback. 
Our goal is to train repair models using unit tests as the correctness signal, while evaluating whether the learned fixer generalizes across realistic sources of bugs.

\paragraph{Code repair.}
Let $x$ denote a programming task, consisting of natural-language instructions, input/output specifications, constraints, and a unit-test suite.
Given a candidate program $c$, running the tests produces a binary verifier
$v(x,c)\in\{0,1\}$, where $v(x,c)=1$ if and only if $c$ passes all tests for task $x$, together with test output $o(x,c)$, such as compilation errors, failed assertions, or stack traces.

We define a valid bug for task $x$ as an executable program $b$ that fails at least one test:
\[
v(x,b)=0.
\]
A repair model, or \emph{fixer}, $\pi_F$ maps the task, buggy program, and test output to a distribution over candidate repairs:
\[
y \sim \pi_F(\cdot \mid x, b, o(x,b)).
\]
A repair succeeds if the repaired program passes the unit tests, i.e., $v(x,y)=1$.
In our main experiments, the fixer outputs a full corrected program rather than a diff.

\paragraph{Evaluation across bug sources.}
In deployment, bugs may come from heterogeneous sources: human programmers, LM coding assistants, or human edits of LM-generated code.
We therefore evaluate repair under a family of bug-source distributions.
For each source $s\in\mathcal{S}$, let $P_s(\cdot\mid x)$ denote the distribution over valid bugs for task $x$.
The repair performance of fixer $\pi_F$ on source $s$ is
\[
\mathrm{Perf}(\pi_F; s)
=
\mathbb{E}_{x \sim \mathcal{D}}
\;
\mathbb{E}_{b \sim P_s(\cdot \mid x)}
\;
\mathbb{E}_{y \sim \pi_F(\cdot \mid x,b,o(x,b))}
\left[
v(x,y)
\right],
\]
where $\mathcal{D}$ is the evaluation distribution over tasks.
In our benchmark, $\mathcal{D}$ is the uniform distribution over a fixed held-out task set, and different bug sources share the same underlying tasks.

We summarize cross-source performance by averaging over bug sources:
\[
\mathrm{Perf}_{\mathrm{avg}}(\pi_F)
=
\frac{1}{|\mathcal{S}|}
\sum_{s\in\mathcal{S}}
\mathrm{Perf}(\pi_F; s).
\]
This criterion evaluates whether a fixer improves broadly across realistic bug sources, rather than only on the distribution induced by a particular bug generator.
We next instantiate $\mathcal{S}$ with \benchmark{}, a controlled benchmark spanning multiple bug sources in LM-assisted programming.

%% file: sections/3.benchmark.tex
\section{\benchmark{}: Controlled Bug-Source Evaluation}
\label{sec:bugbench}

We introduce \benchmark{} to evaluate whether code-repair methods generalize across realistic sources of bugs.
The benchmark is designed to isolate bug-source shift: all sources share the same programming tasks, specifications, and unit tests, and differ only in the buggy implementation provided to the repair model.

\subsection{Benchmark Construction}
\label{sec:bugbench-construction}

We build \benchmark{} from BigCodeBench \citep{zhuo2024bigcodebench}, a code-generation benchmark emphasizing realistic library and API usage.
Each BigCodeBench task contains (i) natural-language programming instructions, (ii) unit tests that define correctness, and (iii) a reference implementation that passes those tests.
We convert each task into a repair instance by preserving the original prompt and unit tests and replacing the reference implementation with a buggy program.

\paragraph{Task structure.} 
A \benchmark{} example consists of programming instructions $x$, a buggy program $b$, and the accompanying unit-test verifier $v(x,\cdot)$.
All buggy programs in \benchmark{} execute under the test harness and fail at least one unit test.
At evaluation, the model receives $(x,b)$ and unit-test feedback $\mathrm{o}(x,b)$ (e.g., failing tests and truncated error traces) and must output a corrected program $y$ that passes all tests.

\paragraph{Bug sources.}
\benchmark{} contains four bug-source variants that reflect common sources of errors in LM-assisted programming.
All variants are constructed on the same underlying tasks; only the buggy implementation differs.

\begin{itemize}
  \setlength{\itemsep}{1pt}
  \setlength{\parskip}{0pt}
  \setlength{\topsep}{2pt}
  \setlength{\partopsep}{0pt}
  \setlength{\parsep}{0pt}

  \item \textbf{\textsc{Human}.}
  Annotators introduce bugs into each task's human-written reference solution.
  They are instructed to make 1--4 localized edits that preserve executability while causing at least one unit test to fail.
  We encourage realistic developer mistakes, such as off-by-one errors, wrong constants, missing edge cases, or API misuse, rather than syntax-breaking edits.

  \item \textbf{\textsc{Human-Edited LM}.}
  To model human-in-the-loop errors, we first prompt \texttt{gpt-5-mini} to solve the task and retain an incorrect program that executes but fails at least one unit test.
  Annotators then edit this draft while keeping it executable and incorrect.
  This source captures mistakes that can arise when developers modify, integrate, or partially correct LM-generated code.

  \item \textbf{\textsc{LM Errors (Qwen-7B)}.}
  We prompt \texttt{Qwen2.5-Coder-7B-Instruct} to solve each task, not to generate bugs, and retain incorrect programs that execute but fail at least one unit test.
  This source captures errors produced by a weaker code LM under standard code-generation prompting.

  \item \textbf{\textsc{LM Errors (gpt-oss-20b)}.}
  We apply the same procedure with \texttt{gpt-oss-20b}, yielding errors from a stronger code model.
\end{itemize}

Together, these variants cover three realistic bug-source families: human-written bugs, human-edited LM bugs, and unedited LM-generated bugs.
Because the task and tests are fixed across sources, \benchmark{} directly measures how repair performance changes with the origin of the bug.

\paragraph{Repair interface.}
We evaluate repair models using a full-program repair interface with unit-test feedback.
That is, the model receives the task description, buggy program, and truncated test output, and returns a complete corrected program.
In \cref{appendix:benchmark,tab:frontier}, we compare this interface with full-program repair without test traces and diff-based patching.
Test feedback improves repair performance, while diff-based repair often underperforms due to brittle formatting and patch application.
We therefore use the test-trace interface in all main experiments.

\paragraph{Benchmark analyses.}
We provide per-source examples in \cref{appendix:benchmark-examples} and full construction details, including filtering criteria and sampling budgets, in \cref{appendix:benchmark_construction}.
In \cref{appendix:benchmark_abla}, we evaluate frontier models on \benchmark{} and show that repair is distinct from code generation: models often solve tasks from scratch that they fail to repair, and vice versa.
We also analyze bug-source structure by categorizing bugs into coarse error types and by measuring $k$NN source purity using \texttt{voyage-code-3} embeddings.
These analyses reveal systematic source-dependent failure patterns and strong within-source clustering.

\benchmark{} lets us test whether synthetic bug--fix training improves repair across realistic bug sources, rather than only on the distribution induced by a particular generator.
We next describe generator--fixer self-play and then introduce \ourslong{}, which anchors self-play to a small reference set to reduce bug-source drift.

%% file: sections/4.method.tex
\section{Self-Play for Code Repair}
\label{sec:analysis}

We now describe the training procedures used to scale code-repair supervision.
We first introduce generator--fixer self-play, which uses unit-test outcomes as rewards.
We then show that this unit-test-only objective can drift toward unrealistic bug distributions, and introduce \ourslong{} to anchor self-play to reference bug sources.

\subsection{Generator--Fixer Self-Play}
\label{sec:standard_self_play}

For each training task, we assume access to the programming specification $x$, unit tests, and a correct reference implementation $c^\star$.
We train a single policy $\pi_\theta$ to play two roles: a \emph{generator} $G$ that edits correct programs into buggy ones, and a \emph{fixer} $F$ that repairs the resulting bugs (\cref{fig:1}).
We denote the corresponding role-conditioned distributions by $\pi_G$ and $\pi_F$.

The generator samples a candidate buggy program
\[
b \sim \pi_G(\cdot \mid x, c^\star),
\]
and we run the unit tests to obtain test output $o(x,b)$.
We call $b$ \emph{valid} if it executes under the test harness and fails at least one unit test.
Invalid bugs receive a penalty and are not used for fixer training.
For valid bugs, the fixer samples one or more candidate repairs conditioned on the task, buggy program, and test output:
\[
y \sim \pi_F(\cdot \mid x,b,o(x,b)).
\]

\subsection{Correctness and Difficulty Rewards}
\label{sec:base_rewards}

\paragraph{Fixer reward.}
The fixer is rewarded for producing repairs that pass the unit tests.
For a candidate repair $y$, we use
\[
r^{\textsc{F}}(x,b,y)=v(x,y),
\]
where $v(x,y)=1$ if and only if $y$ passes all tests for task $x$.

\paragraph{Generator reward.}
The generator should produce bugs that are valid, challenging, and still solvable by the current fixer.
A single repair attempt gives a noisy estimate of difficulty, so we estimate the fix rate of a generated bug using $K$ independent repair attempts.
For a valid bug $(x,b)$, we sample
\[
y^{(1)},\ldots,y^{(K)} \sim \pi_F(\cdot \mid x,b,o(x,b))
\]
and define
\[
\rho(x,b)=\frac{1}{K}\sum_{k=1}^{K} v(x,y^{(k)}).
\]
Bugs with $\rho(x,b)=1$ are too easy, while bugs with $\rho(x,b)=0$ provide little useful training signal because the current fixer cannot repair them.
We therefore reward valid bugs whose fix rate falls in an intermediate difficulty band:
\[
r^{\textsc{G}}_{\mathrm{base}}(x,b)=
\begin{cases}
-1, & b \text{ is invalid},\\
1, & \rho(x,b)\in[\rho_\ell,\rho_h],\\
-\alpha, & \rho(x,b)\in\{0,1\},\\
0, & \text{otherwise.}
\end{cases}
\]

\subsection{Optimization}
\label{sec:opt}

We optimize the generator--fixer loop with GRPO.
For each task $x$, we sample $G=4$ candidate bugs $b_i \sim \pi_G(\cdot\mid x,c^\star)$.
For each valid bug $(x,b_i)$, we sample $K=4$ independent repair attempts
\[
y_i^{(1:K)} \sim \pi_F(\cdot\mid x,b_i,o(x,b_i))
\]
and compute the empirical fix rate
\[
\rho(x,b_i)=\frac{1}{K}\sum_{k=1}^{K} v(x,y_i^{(k)}).
\]

Let $R^{\textsc{G}}(x,b)$ denote the generator reward and $R^{\textsc{F}}(x,b,y)$ denote the fixer reward.
In standard self-play, $R^{\textsc{F}}$ is the unit-test pass indicator and $R^{\textsc{G}}$ is the difficulty-shaped reward from \cref{sec:base_rewards}.
\ours{} modifies this objective by adding a reference-similarity term to $R^{\textsc{G}}$ and by mixing reference bugs into fixer training (\cref{sec:method-anchoring}).

\paragraph{Generator update.}
For each task $x$, we compute group-normalized advantages over the $G$ sampled bugs:
\begin{align*}
\hat{A}^{\textsc{G}}_i
&=
\frac{
R^{\textsc{G}}(x,b_i)-\mu^{\textsc{G}}(x)
}{
\sigma^{\textsc{G}}(x)+\epsilon
},
\\
\mu^{\textsc{G}}(x),\sigma^{\textsc{G}}(x)
&\text{ computed over } i\in\{1,\dots,G\}.
\end{align*}
We then update $\pi_G(\cdot\mid x,c^\star)$ using the clipped GRPO objective.

\paragraph{Fixer update.}
For each valid generated bug $b_i$, we compute group-normalized advantages over the $K$ repair attempts:
\begin{align*}
\hat{A}^{\textsc{F}}_{i,k}
&=
\frac{
R^{\textsc{F}}(x,b_i,y_i^{(k)})-\mu^{\textsc{F}}(x,b_i)
}{
\sigma^{\textsc{F}}(x,b_i)+\epsilon
},
\\
\mu^{\textsc{F}}(x,b_i),\sigma^{\textsc{F}}(x,b_i)
&\text{ computed over } k\in\{1,\dots,K\}.
\end{align*}
We update $\pi_F(\cdot\mid x,b_i,o(x,b_i))$ with the same clipped GRPO objective, computing the loss only on tokens generated by the policy and masking prompt tokens.

\subsection{Distribution Drift Under Standard Self-Play}
\label{sec:drift}

The rewards above encourage bugs that are valid and appropriately difficult, but they do not encourage bugs to be realistic.
Unit tests can identify whether a candidate bug changes program behavior, but they do not indicate whether the edit resembles a bug written by a human programmer or produced by an LM coding assistant.
As a result, the generator can drift toward idiosyncratic test-failing edits that remain useful under the self-play reward but differ from realistic bug sources.

\begin{figure*}[t]
  \centering
  \begin{subfigure}[t]{0.44\textwidth}
    \centering
    \includegraphics[width=\linewidth]{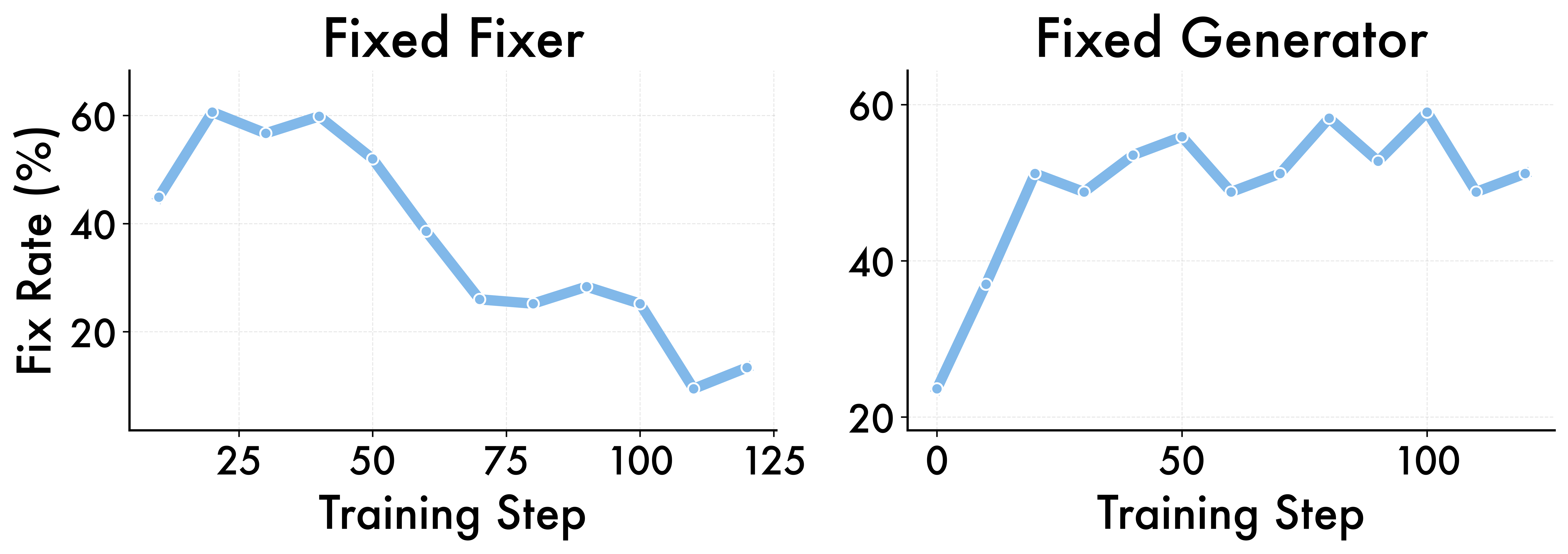}
    \caption{
    \textbf{Co-evolution of standard self-play.}
    With a fixed fixer checkpoint, fix rate declines over generator training, indicating that the generator produces harder bugs.
    With a fixed generator checkpoint, fix rate increases as the fixer trains, indicating improved repair on generated bugs.
    }
    \label{fig:selfplay_training_dynamics}
  \end{subfigure}\hfill
  \begin{subfigure}[t]{0.54\textwidth}
    \centering
    \includegraphics[width=\linewidth]{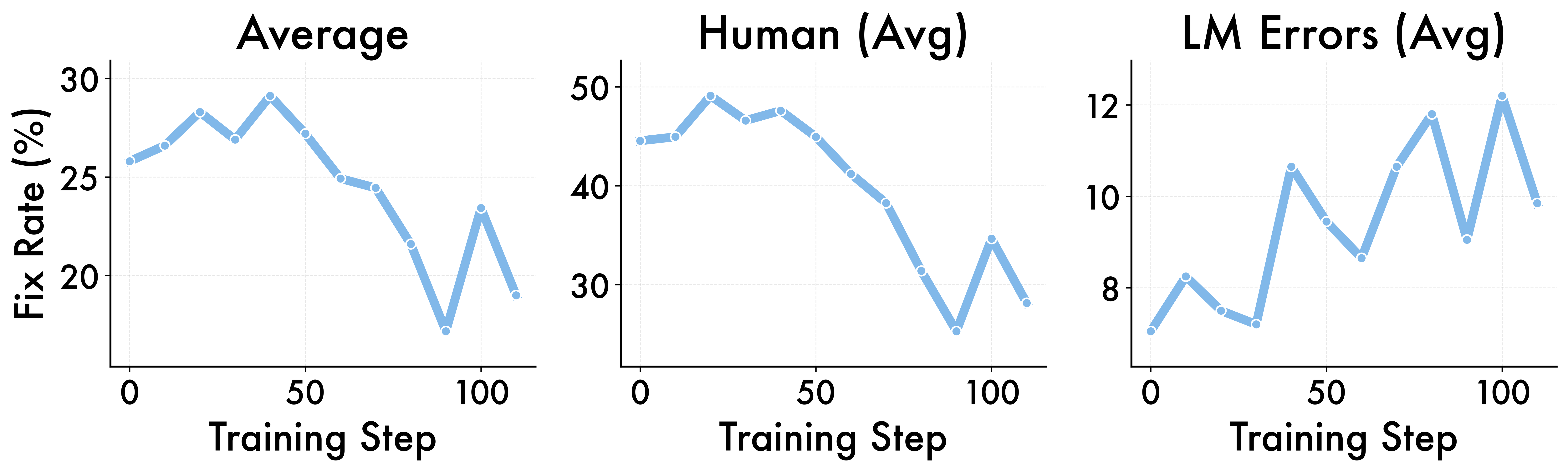}
    \caption{
    \textbf{Standard self-play exhibits distribution drift.}
    Fix rate improves early but later regresses, most strongly on human-originated bugs.
    This suggests overfitting to the generator's shifting synthetic bug distribution.
    }
    \label{fig:standard_selfplay}
  \end{subfigure}
  \caption{\textbf{Self-play training dynamics and distribution drift.}}
  \label{fig:selfplay_training_curves}
\end{figure*}

\Cref{fig:selfplay_training_dynamics} shows that standard self-play produces the intended co-evolutionary dynamics: as the generator trains, its bugs become harder for a fixed fixer, and as the fixer trains, it improves on bugs from a fixed generator.
However, \cref{fig:standard_selfplay} shows that these gains do not reliably transfer to realistic bug sources.
Performance improves early but later degrades, especially on human-originated bugs.
This pattern is consistent with distribution drift: the fixer improves on the generator's evolving synthetic distribution while losing robustness to target bug sources.
We therefore modify the self-play objective to anchor generation and fixer training to reference bug distributions.

\subsection{Anchoring Self-Play to Reference Bug Sources}
\label{sec:method-anchoring}

\ourslong{} assumes access to a small reference set of valid bugs, $\mathcal{D}_{\mathrm{ref}}$, drawn from the training tasks and disjoint from evaluation.
The reference set provides a weak realism signal for self-play.
We use it in two complementary ways:
(i) \emph{reference mixing}, which exposes the fixer to reference-source bugs during training, and
(ii) \emph{similarity-guided shaping}, which nudges the generator toward reference-like edits.

\subsubsection{Reference Mixing for Fixer Training}
\label{sec:reference_mix}

For a training task with an associated reference bug $b^{\mathrm{ref}}\in\mathcal{D}_{\mathrm{ref}}$, we replace the generated bug with probability $p_{\mathrm{mix}}$ and train the fixer on $(x,b^{\mathrm{ref}},o(x,b^{\mathrm{ref}})).$
On these mixed episodes, we update only the fixer.
We do not update the generator because the reference bug was not sampled from $\pi_G$; empirically, updating the generator on mixed episodes reduced performance (\cref{tab:gen_on_mixed}).

\subsubsection{Similarity-Guided Generator Shaping}
\label{sec:sim_reward}

Reference mixing improves fixer training but does not directly change the generator's distribution.
To guide generation toward realistic bugs, we add an auxiliary reward based on similarity to the reference set.

\paragraph{Edit embeddings.}
For each generated bug $b$, we compute the unified diff between the reference solution $c^\star$ and $b$.
We embed this diff with a code embedding model, \texttt{voyage-code-3}, and compute its average cosine similarity to the $k$ nearest reference edit embeddings.
We map the resulting score to $[0,1]$ and denote it by $\mathrm{sim}_{01}(b)$.

\paragraph{Centered similarity reward.}
The similarity reward should act as a shaping term: it should bias the generator toward reference-like edits without overwhelming the unit-test reward.
Because the absolute scale of embedding similarities can vary across batches and embedding models, we subtract a running baseline.
At training step $t$, we maintain
\[
B_t \leftarrow \beta B_{t-1} + (1-\beta)\,
\mathbb{E}_{b \sim \mathrm{batch}}\!\left[\mathrm{sim}_{01}(b)\right],
\]
and define
\[
\delta_t(b)=\mathrm{sim}_{01}(b)-B_t.
\]
Thus, $\delta_t(b)$ measures whether a generated bug is more or less reference-like than the generator's current average output.
For valid generated bugs, the final generator reward is
\[
r^{\textsc{G}}(x,b)
=
r^{\textsc{G}}_{\mathrm{base}}(x,b)
+
\lambda\,\operatorname{clip}\!\left(\delta_t(b)\right),
\]
where $\operatorname{clip}(\cdot)$ clips the shaping term to a fixed range.
This centered reward is more stable than using an uncentered similarity score, since it is less sensitive to shifts in the absolute magnitude of embedding similarities.
Hyperparameters are given in \cref{appendix:hyperparameters}.

%% file: sections/5.experiments.tex
\section{Experimental Setup}
\label{sec:experimental_setup}

\begin{figure*}[t]
  \centering
  \includegraphics[width=0.78\textwidth]{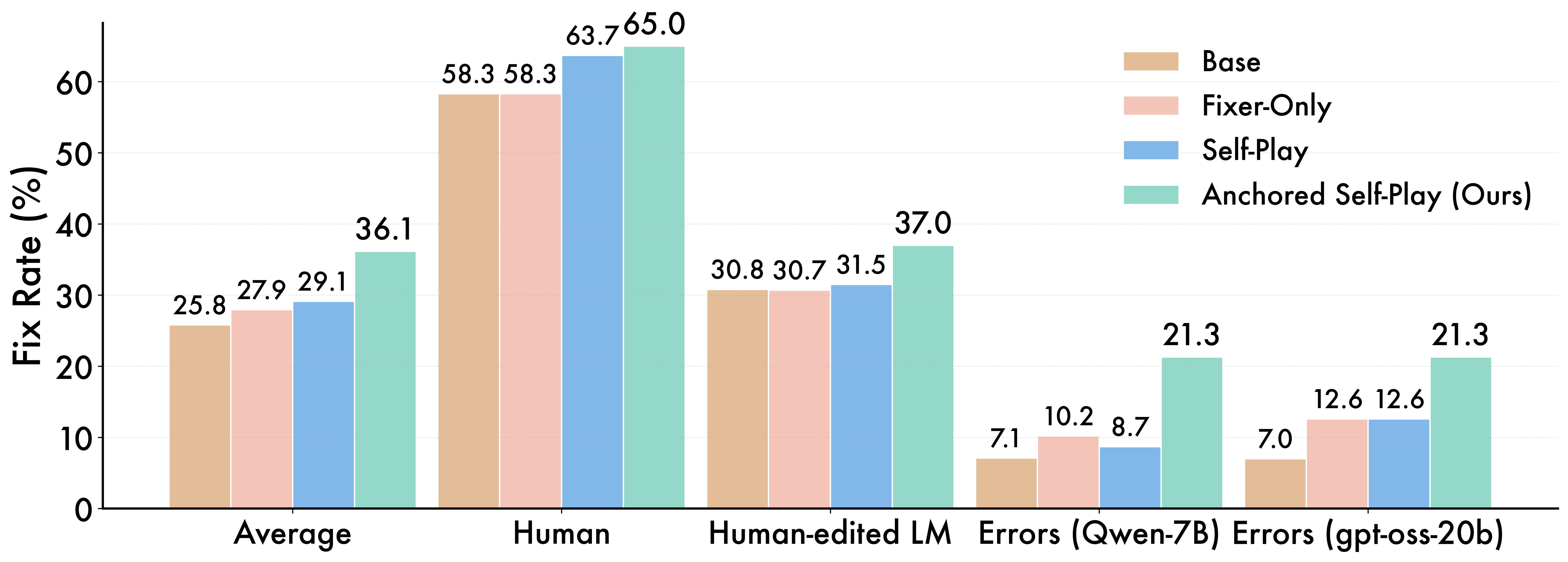}
  \caption{
    \textbf{Fix rates across bug sources on \benchmark{}.}
    Standard self-play yields inconsistent gains across bug sources, consistent with distribution drift.
    \ours{} improves repair across every source and achieves the best overall fix rate, outperforming standard self-play by $+7.0$ pp / $24\%$ relative.
    Gains hold for both LM-generated bugs ($+11$ pp / $100\%$ relative) and human-originated bugs ($+3.4$ pp / $7.1\%$ relative).
  }
  \label{fig:anchored}
\end{figure*}

\begin{figure*}[t]
  \centering
  \includegraphics[width=0.96\textwidth]{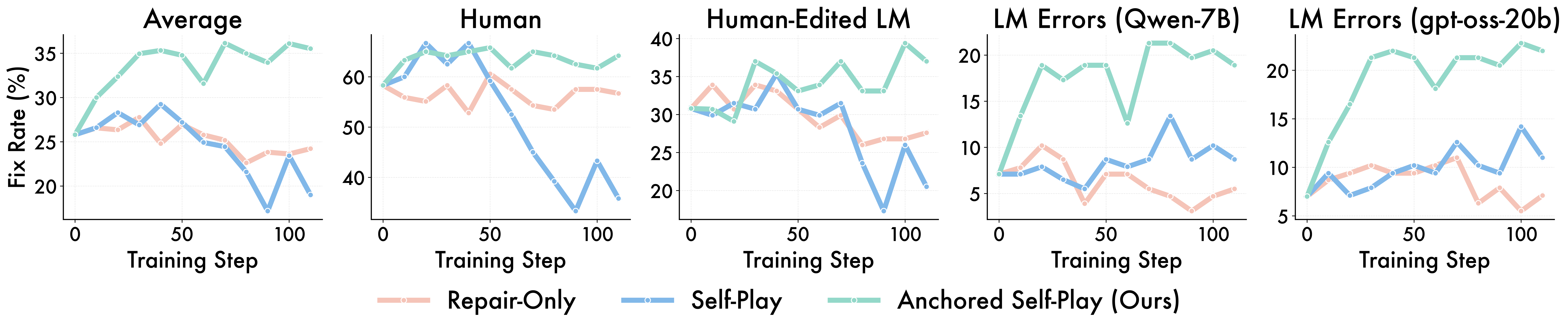}
  \caption{
  \textbf{Anchoring stabilizes self-play and improves cross-source repair.}
  We plot fix rate versus training step for Fixer-Only, standard Self-Play, and \ours{} on held-out bugs from each \benchmark{} source and their average.
  Standard self-play improves early but later degrades on realistic bug sources, especially Human and Human-Edited LM bugs.
  In contrast, \ours{} yields higher and more stable performance across sources.
  }
  \label{fig:selfplay_learning_curves}
\end{figure*}

\paragraph{Data and splits.}
We train on $900$ BigCodeBench tasks \citep{zhuo2024bigcodebench}.
For evaluation, we use $127$ held-out tasks that appear in every \benchmark{} split.
Thus, all bug sources share the same underlying tasks, specifications, and unit tests; only the buggy implementation differs.
For validation and checkpoint selection, we use $81$ additional tasks from all \benchmark{} splits, disjoint from both train and test.
We report results on four bug sources: \benchmark{}\textsc{-Human}, \benchmark{}\textsc{-Human-Edited LM}, \benchmark{}\textsc{-Qwen-7B}, and \benchmark{}\textsc{-gpt-oss-20b}.

\paragraph{Reference pool.}
\ours{} uses a reference pool of $900$ bugs sampled from the training splits of \benchmark{}, with equal representation from the four bug sources.
This pool serves two purposes: it provides reference edit embeddings for the generator's similarity reward, and it supplies reference bugs for fixer training through reference mixing.
For parity, the Fixer-Only baseline also mixes the same reference bugs into fixer training, but does not use the similarity-based generator reward.

\paragraph{Initialization.}
Unless otherwise stated, both generator and fixer are initialized from \texttt{Qwen2.5-Coder-7B-Instruct} \citep{hui2024qwen2}.
The two roles share a single set of weights and are trained with GRPO.

\paragraph{Comparisons.}
We compare \ours{} against three main baselines:
\begin{itemize}[leftmargin=*, nosep]
  \item \textbf{Base Model:} the pretrained model used directly as a fixer, without post-training.
  \item \textbf{Fixer-Only:} a frozen base-model generator samples bugs, while only the fixer is trained with GRPO. Reference bugs are mixed into fixer batches for parity with \ours{}.
  \item \textbf{Self-Play:} the generator and fixer are trained jointly with GRPO using the unit-test-only self-play objective.
\end{itemize}

\paragraph{Training and evaluation protocol.}
For each training task, the generator samples candidate buggy programs.
We retain programs that execute under the test harness and fail at least one unit test.
The fixer receives the task description, buggy code, and truncated unit-test output, and is rewarded for producing a program that passes the tests.
During training, the fixer samples up to $K=4$ repairs per bug.
At test time, we use a single repair attempt per bug and decode greedily with temperature $0.0$.
Full hyperparameters and prompts are provided in \cref{sec:appendix-training-details}; additional baselines are reported in \cref{appendix:codegengrpo_c2}.

\section{Main Results}
\label{sec:main_results}

\Cref{fig:anchored} compares \ours{} with standard self-play and fixer-only training on \benchmark{}.
\ours{} achieves the best fix rate on every bug source, improving average fix rate by $24\%$ relative / $+7.0$ pp absolute over standard self-play and by $29\%$ relative / $+8.2$ pp absolute over Fixer-Only.
The largest gains are on LM-generated bugs: $+145\%$ relative / $+12.6$ pp absolute on \textsc{Qwen-7B} bugs, and $+69\%$ relative / $+8.7$ pp absolute on \textsc{gpt-oss-20b} bugs.
Anchoring also improves both Human and Human-Edited LM splits relative to standard self-play, with gains of $+17.5\%$ relative / $+5.5$ pp absolute on Human-Edited LM and $+2.0\%$ relative / $+1.3$ pp absolute on Human.

Standard self-play improves the average fix rate over Fixer-Only by $+1.2$ pp, but its gains are not consistent across sources: it improves some splits while degrading on \textsc{Qwen-7B} bugs by $-1.5$ pp.
In contrast, \ours{} improves all sources simultaneously.
The Base Model and Fixer-Only have similar average performance, but Fixer-Only shifts performance toward LM-generated bugs: it improves \textsc{Qwen-7B} from $7.1\%$ to $10.2\%$ and \textsc{gpt-oss-20b} from $7.0\%$ to $12.6\%$, while leaving Human performance essentially unchanged ($30.8\%$ vs.\ $30.7\%$).
These results show why evaluating only one bug source can be misleading: training can improve one source while failing to improve, or even hurting, another.

We provide pass@$k$ results in \cref{appendix:passk}.
We also show that \ours{} can improve larger 30B+-parameter code models when used as a test-time fixer in \cref{appendix:testtime_fixer}.

\paragraph{\ours{} produces more reference-like bugs.}
We use embedding-based diagnostics to verify that the similarity reward changes the generator's bug distribution.
As shown in \cref{fig:similarity_and_buckets}, the mean $k$-NN similarity between generated bugs and the reference pool increases under \ours{}, indicating that similarity shaping moves generations toward the target bug sources.
We also stratify benchmark performance by generation-similarity quantiles and find that \ours{} outperforms standard self-play within each bin.
Thus, the gains are not explained solely by generating higher-similarity bugs; anchoring also improves the fixer's robustness within comparable similarity regimes.
Additional analyses show that \ours{} improves consistently across semantic bug categories (\cref{appendix:bugtype_eval}), and qualitative examples are provided in \cref{fig:example-bugs} and \cref{appendix:additional_c}.

\section{Ablations}
\label{sec:ablations}

\subsection{Anchoring Components}

Table~\ref{tab:components_ablation} ablates the two components of \ours{}: reference mixing for fixer training and the embedding-similarity reward for bug generation.
Starting from standard self-play, each component alone improves average fix rate modestly.
Combining them yields a substantially larger gain, suggesting that the two components address complementary failure modes.
The similarity reward steers the generator toward reference-like bugs, while reference mixing keeps the fixer exposed to realistic bugs as the generator's distribution evolves.

\begin{table}[t]
\centering
\footnotesize
\captionsetup{font=footnotesize}

\begin{subtable}{\columnwidth}
\centering
\caption{
\textbf{Ablation of anchoring components.}
Reference mixing exposes the fixer to reference bugs during training, while similarity shaping adds an embedding-based reward that guides the generator toward reference-like edits.
The full \ours{} method combines both and achieves the highest average fix rate across \benchmark{} splits.
}
\label{tab:components_ablation}
\vspace{2pt}
\setlength{\tabcolsep}{3pt}
\begin{tabular}{lccc}
\toprule
\textbf{Method} & \textbf{Ref. Mix} & \textbf{Sim. Reward} & \textbf{Fix (\%)} \\
\midrule
Base Model & & & 25.8 \\
Self-Play & & & 29.1 \\
 + Reference Bug Mixing  & \checkmark & & 29.5 \\
 + Similarity Shaping  &  & \checkmark & 30.9 \\
\textbf{\ours{}} (Ours)  & \checkmark & \checkmark & \textbf{36.1} \\
\bottomrule
\end{tabular}
\end{subtable}

\vspace{4pt}

\begin{subtable}{\columnwidth}
\centering
\caption{
\textbf{Effect of reference-pool composition.}
We vary which bug sources are included in the reference pool.
Human-only references yield the strongest performance on Human bugs, while LM-only references shift improvements toward LM-originated bugs.
The mixed reference pool used by \ours{} gives the best overall performance and the strongest gains on LM-generated sources, illustrating that the reference pool steers which bug patterns self-play emphasizes.
}
\label{tab:ref_pool_composition}
\vspace{2pt}
\setlength{\tabcolsep}{3pt}
\resizebox{\columnwidth}{!}{%
\begin{tabular}{lccccc}
\toprule
 & \multicolumn{5}{c}{\textbf{Fix Rate (\%)}} \\
\cmidrule(lr){2-6}
\textbf{Ref. Pool} & \textbf{Overall} & \textbf{Human} & \textbf{Hum.-Ed.} & \textbf{Qwen} & \textbf{GPT-OSS} \\
\midrule
Human-only  &  34.4 & \textbf{67.5} &  37.8 &  11.8 & 11.0 \\
LM-only     &  32.0 & 65.4 &  \textbf{41.7} &  13.4 & 17.3 \\
Self-Play    & 29.1 & 63.7 & 31.5 & 8.7 & 12.6 \\
\textbf{\ours{} (Ours)} & \textbf{36.1} & 65.0 & 37.0 & \textbf{21.3} & \textbf{21.3} \\
\bottomrule
\end{tabular}%
}
\end{subtable}
\end{table}

\subsection{Reference-Pool Composition}

Table~\ref{tab:ref_pool_composition} varies the sources used to construct the reference pool.
The choice of reference bugs affects which bug patterns self-play emphasizes.
Human-only references give the highest fix rate on Human bugs, while LM-only references improve LM-originated bugs relative to standard self-play and give the highest score on Human-Edited LM.
The mixed pool used by \ours{} achieves the best overall fix rate and the strongest performance on both LM-generated sources.
These results suggest that the reference pool can steer anchoring toward particular deployment sources, while a diverse pool provides the best overall performance.

\subsection{Reference Set Size}

\begin{figure}[t]
  \centering
  \includegraphics[width=\linewidth]{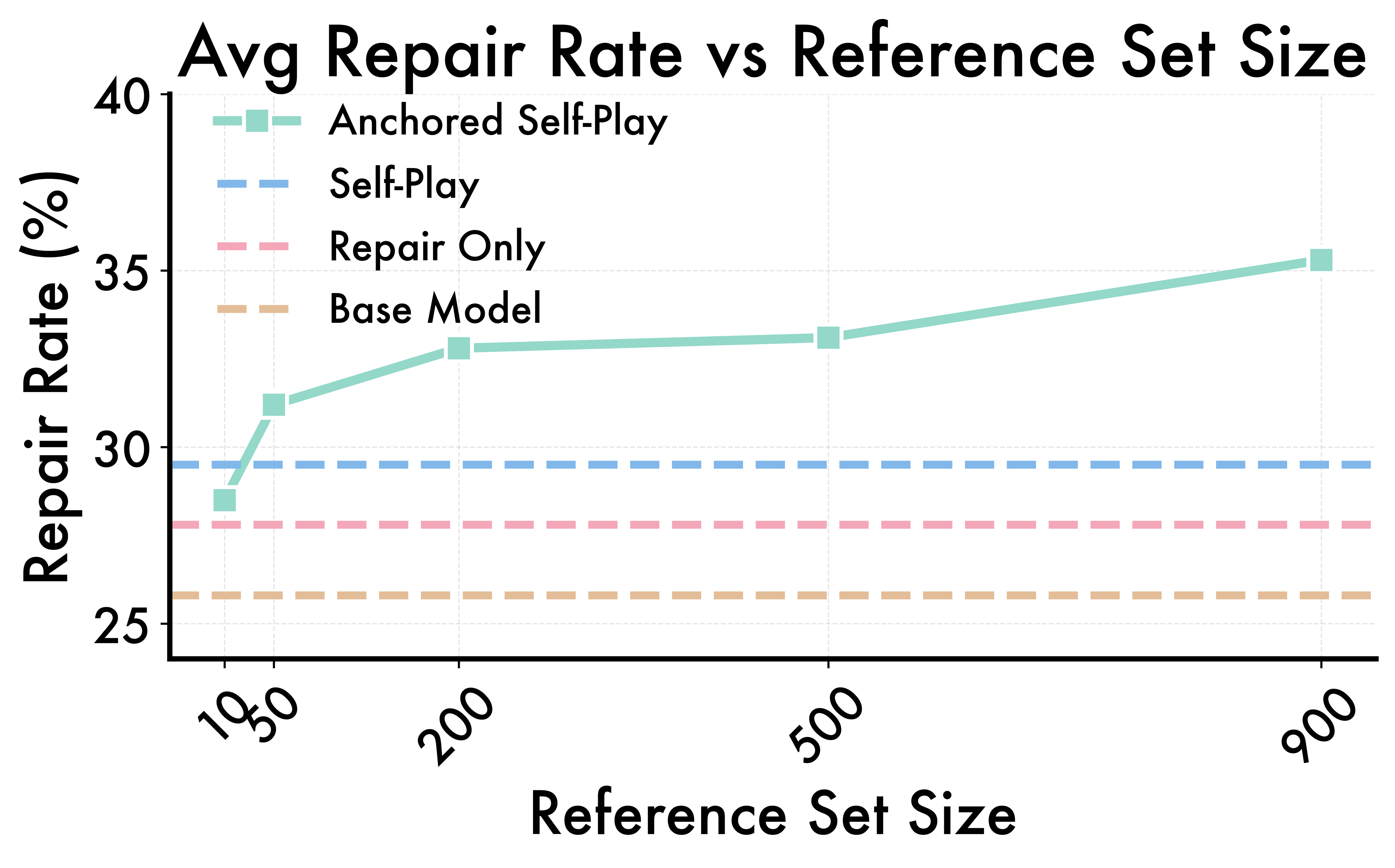}
  \caption{
  \textbf{Reference set scaling.}
  We vary the number of reference bugs used for anchoring and report fix rates on each \benchmark{} split and their average.
  \ours{} improves with small reference sets and continues to benefit from larger pools, with the largest gains on LM-generated bug sources.
  Human and Human-Edited LM performance remain strong across reference-set sizes.
  }
  \label{fig:reference-size}
\end{figure}

\Cref{fig:reference-size} varies the number of reference bugs used for anchoring, sampled uniformly across Human, Human-Edited LM, \textsc{Qwen-7B}, and \textsc{gpt-oss-20b} sources.
\ours{} improves with as few as $50$ reference examples, indicating that anchoring is sample-efficient.
Performance continues to improve as the reference set grows, suggesting that larger pools provide broader coverage of realistic bug patterns.

Additional robustness ablations in \cref{appendix:additional_c} and \cref{appendix:codegengrpo_c2} show that \ours{}'s gains are robust across embedding models and $k$-NN pooling parameters (\cref{tab:knn_k}), base models (\cref{tab:alt_base}), shared versus decoupled generator/fixer weights (\cref{tab:decoupled}), and task distributions (\cref{tab:debugbench_results}).

%% file: sections/6.related_work.tex
\section{Related Work}

\paragraph{Code repair and bug-source variation.}
Automated program repair is commonly evaluated on curated real-world bug datasets such as Defects4J, BugsInPy, ManyBugs, and Bears, as well as short unit-testable benchmarks such as QuixBugs \citep{just2014defects4j,widyasari2020bugsinpy,leGoues2015manybugs,madeiral2019bears,lin2017quixbugs}.
Related code-generation benchmarks, including HumanEval, MBPP, APPS, EvalPlus, and LiveCodeBench, show that fluent model outputs often fail functional tests \citep{chen2021evaluating,austin2021program,hendrycks2021apps,liu2023evalplus,jain2024livecodebench}, while repository-level benchmarks such as SWE-bench evaluate end-to-end issue resolution with realistic context and tooling \citep{jimenez2023swe,yang2025swe,pham2025swe}.
Across these settings, performance is sensitive to the bug distribution: human-written bugs, synthetic mutations, and errors in model-generated code induce different failure modes \citep{he2022distribution,xu2022systematic,sonwane2025bugpilotcomplexbuggeneration,dou2025whatswrongcodegenerated,yang2025coastenhancingcodedebugging}.
\benchmark{} complements prior benchmarks by holding the task, specification, and tests fixed while varying only the bug source, enabling controlled evaluation of cross-source generalization.

\paragraph{Synthetic bug generation.}
Prior work scales repair supervision by generating synthetic bugs, often under constraints that preserve realism.
BugLab uses predefined mutation operators for self-supervised bug localization and repair \citep{allamanis2021buglab,forrest2009genprog}; repository-level systems use grounding from repository structure, tests, issue context, or patch provenance \citep{yang2025swe,pham2025swe,wei2025ssr,10.1145/3551349.3556926,10.1145/3631972}; and Break-it-Fix-it trains paired bug introducers and fixers using a critic such as a parser or compiler \citep{yasunaga2021breakitfixit,long2016prophet,chen2018sequencer}.
In contrast, we study open-ended bug generation, where the generator can apply arbitrary text edits and unit tests verify failure but only weakly constrain realism.
Our method adds a soft realism signal through embedding similarity to reference bugs while preserving open-ended generation.

\paragraph{Self-play and curricula.}
Self-play has been used to generate curricula by proposing tasks near a learner's frontier and learning from rollouts \citep{bengio2009curriculum,silver2017mastering,cheng2024spar,kuba2025languageselfplay,chen2024self,zhao2025absolutezero,huang2025rzero}.
In language and code domains, related work uses self-generated data for reasoning, theorem proving, adaptive testing, open-ended task design, and proposer--solver frameworks for synthesis, testing, or verification \citep{poesia2024theorem,dong2025theorem,chen2025selfquestioning,liu2025spice,yu2025guided,ribeiro2022adatest,colas2022autotelic,parkerholder2022regretbased,teodorescu2023codeplay,pourcel2024aces,lin2025solvver,wang2025cure,wilf2025psv}.
We apply this curriculum perspective to code repair and identify a failure mode of unit-test-only self-play: without an explicit realism signal, the generator can produce difficult test-failing edits that do not resemble realistic bugs.

%% file: sections/7.conclusion.tex
\section{Discussion}

We study whether unit tests can scale code-repair supervision through open-ended synthetic data generation.
Generator--fixer self-play creates an adaptive curriculum by training a model to generate bugs that fail tests and repair bugs to pass them.
However, we find that unit-test-only self-play can drift toward valid but unrealistic bugs, improving on its own generated distribution while weakening generalization to realistic bug sources.

\ourslong{} mitigates this drift by anchoring self-play to a small reference set, using an embedding-similarity reward for bug generation and reference mixing for fixer training.
Using \benchmark{}, a controlled benchmark spanning human-written, human-edited LM, and LM-generated bugs, we show that \ours{} improves average repair over standard self-play and reduces bug-source drift.
These results highlight the need for explicit realism signals in unit-test-only self-play.
Future work could explore stronger realism objectives, such as learned bug-style critics, preference-based rewards, or richer anchoring signals beyond code embeddings.

\paragraph{Limitations.}
Our experiments focus on Python function-level repair with unit-test feedback.
This controlled setting isolates bug-source shift, but does not capture repository-level challenges such as multi-file fault localization, dependency management, build-system interaction, or long-horizon tool use.
Our realism signal also relies on a frozen code-embedding model and a finite reference pool; learned critics or human preference data may provide stronger signals in future work.

\section*{Impact Statement}

This work aims to improve the reliability of language-model-based code repair.
By studying generalization across bug sources, our results can help build automated programming tools that are more robust to realistic errors.
As with other automated software-engineering systems, generated fixes should be validated before deployment.
We do not anticipate additional societal risks beyond those associated with broader use of machine learning in software engineering.

%% file: sections/8.appendix.tex
\clearpage
\onecolumn
\appendix

\section{\benchmark{}}
\label{appendix:benchmark}

\subsection{Examples}
\label{appendix:benchmark-examples}

\begin{figure}[h]
\centering
\begin{codebox}[left=5pt,right=5pt]{BugSourceBench TaskID 1063}
\fontsize{7}{8}\selectfont
Performs Principal Component Analysis (PCA) on the sum of rows of a 2D numpy array and plots the explained variance ratio.\\
\textit{Note that:} the title of the plot is set to "Explained Variance Ratio of Principal Components". The function assumes that 'arr' is a valid 2D numpy array. Only the first principal component is considered in this analysis. The plot illustrates the proportion of the dataset's variance that lies along the axis of this first principal component.\\
\textit{The function should output with:}
ax (matplotlib.axes.Axes) -- an Axes object from matplotlib.
\end{codebox}

\vspace{-6pt}

\begin{minipage}[t]{0.498\textwidth}
\begin{codebox}{Human}
\begin{lstlisting}[style=py]
import numpy as np
from matplotlib import pyplot as plt
from sklearn.decomposition import PCA
def task_func(arr):

    row_sums = arr.sum(axis=1)
    
    @@pca = PCA(n_components=10)@@ 
    pca.fit(row_sums.reshape(-1, 1))

    # Plotting (requires matplotlib and sklearn)

    _, ax = plt.subplots()
    
    @@ax.bar([0], pca.explained_variance_ratio)@@
    ax.set_title("Explained Variance Ratio of Principal Components")
    ax.set_xticks([0])
    ax.set_xticklabels(["PC1"])

    return ax
    
\end{lstlisting}
\end{codebox}
\end{minipage}\hfill
\begin{minipage}[t]{0.498\textwidth}
\begin{codebox}{Human-Edited LM}
\begin{lstlisting}[style=py]
import numpy as np
from matplotlib import pyplot as plt
from sklearn.decomposition import PCA
def task_func(arr):
    if arr.size == 0:
        raise ValueError("Input array is empty.")

    @@summed = np.sum(arr, axis=1)@@
    X = summed.reshape(-1, 1)
    pca = PCA(n_components=1)
    pca.fit(X)

    fig, ax = plt.subplots()
    @@ax.bar([1], pca.explained_variance_, color="tab:blue")@@
    ax.set_xticks([1])
    ax.set_xticklabels(["PC1"])
    ax.set_ylabel("Explained Variance Ratio")
    ax.set_title("Explained Variance Ratio of Principal Components")

    return ax
\end{lstlisting}
\end{codebox}
\end{minipage}
\vspace{1pt}
\begin{minipage}[t]{0.498\textwidth}
\begin{codebox}{LM Errors (Qwen-7B)}
\begin{lstlisting}[style=py]
import numpy as np
from matplotlib import pyplot as plt
from sklearn.decomposition import PCA
def task_func(arr):
    # Ensure arr is a 2D numpy array
    if len(arr.shape) != 2:
        raise ValueError("Input must be a 2D numpy array")
    
    # Sum the rows of the array
    @@summed_data = np.sum(arr, axis=0)@@
    
    # Reshape the data to be 2D for PCA
    @@reshaped_data = summed_data.reshape(1, -1)@@
    
    # Perform PCA
    pca = PCA(n_components=1)
    pca.fit(reshaped_data)
    
    # Get the explained variance ratio
    explained_variance_ratio = pca.explained_variance_ratio_
    
    # Create a plot
    fig, ax = plt.subplots()
    ax.bar(range(1), explained_variance_ratio, tick_label=['PC1'])
    ax.set_title('Explained Variance Ratio of Principal Components')
    
    return ax
\end{lstlisting}
\end{codebox}
\end{minipage}\hfill
\begin{minipage}[t]{0.498\textwidth}
\begin{codebox}{LM Errors (gpt-oss-20b)}
\begin{lstlisting}[style=py]
import numpy as np
from matplotlib import pyplot as plt
from sklearn.decomposition import PCA
def task_func(arr):
    # Sum over rows to get column totals
    @@row_sums = arr.sum(axis=0)@@

    # Reshape to 2D array for PCA (n_samples=1, n_features=number of columns)
    @@X = row_sums.reshape(1, -1)@@

    # Perform PCA with a single component
    pca = PCA(n_components=1)
    pca.fit(X)

    # Extract explained variance ratio for the first component
    evr = pca.explained_variance_ratio_[0]

    # Plot the explained variance ratio
    fig, ax = plt.subplots()
    ax.bar([1], [evr], width=0.5, color='skyblue')
    ax.set_title("Explained Variance Ratio of Principal Components")
    ax.set_xlabel("Principal Component")
    ax.set_ylabel("Explained Variance Ratio")
    @@ax.set_xticks([1])@@
    ax.set_ylim(0, 1)

    return ax
\end{lstlisting}
\end{codebox}
\end{minipage}

\vspace{1pt}
\caption{\benchmark{} example showing buggy programs from different sources for the same task $x$.
The bug patterns differ qualitatively across sources.
The human-written bug changes PCA parameters and plotting behavior; the human-edited LM bug uses explained variance rather than explained-variance ratio and changes the plotted index; Qwen-7B and gpt-oss-20b both make model-like errors involving the aggregation axis and PCA input shape.
These examples illustrate why evaluating a single bug source can miss source-specific repair failures.}
\label{fig:benchmark-examples}
\end{figure}

\subsection{Construction}
\label{appendix:benchmark_construction}

\paragraph{Overview.}
\benchmark{} is built from $1{,}114$ BigCodeBench-style programming tasks.
Each task provides natural-language instructions, unit tests that define correctness, and a reference implementation that passes those tests.
A \benchmark{} instance pairs a task with a buggy program that executes under the test harness and fails at least one unit test.
All \benchmark{} variants share the same underlying tasks and differ only in how the buggy program is produced, enabling controlled comparisons across bug sources.

\paragraph{Bug validity criteria.}
We first remove $26$ BigCodeBench tasks whose reference solutions do not pass the provided unit tests, leaving $1{,}114$ tasks for bug construction.
We accept a candidate program as a valid bug if it executes under the test harness and fails at least one unit test.
We reject candidates that fail due to interpreter or runtime errors identified by pattern matching over test output, such as \texttt{SyntaxError}, \texttt{ImportError}, or \texttt{NameError}.
Assertion failures from the unit tests are treated as valid test failures.

\paragraph{Task structure.}
All datasets are converted to a common \benchmark{} schema.
We store function bodies only, with 4-space indentation, for both \texttt{buggy} and \texttt{canonical\_solution}.
Each example contains:
\begin{itemize}[leftmargin=*, nosep]
    \item \texttt{task\_id}: identifier for the underlying task, shared across bug sources.
    \item \texttt{instruct\_prompt}: natural-language problem statement, including input/output specifications and constraints.
    \item \texttt{buggy}: buggy function body that executes but fails at least one unit test.
    \item \texttt{canonical\_solution}: reference correct function body, used for dataset construction and evaluation but never provided to the fixer.
    \item \texttt{test}: unit-test harness used to evaluate candidate repairs.
    \item \texttt{complete\_prompt}: full prompt used in the default repair interface, including instructions, buggy code, and test feedback.
    \item \texttt{code\_prompt}: code-only prompt segment containing the function context.
    \item \texttt{entry\_point}: function name called by the test harness.
    \item \texttt{doc\_struct}: structured metadata extracted from the problem statement, when available.
    \item \texttt{libs}: libraries or modules required by the task.
    \item \texttt{test\_output}: truncated feedback from running \texttt{buggy} on \texttt{test}, used for analysis and as repair context.
\end{itemize}
This schema lets us swap bug sources while holding the task, specification, and tests fixed.

\paragraph{Bug sources.}
We construct four bug-source variants, each providing a different buggy program for the same underlying BigCodeBench tasks.
For fair comparison, we retain only task IDs that appear in all variants.

\begin{itemize}[leftmargin=*, nosep]
  \item \textbf{\benchmark{}-\textsc{Human}.}
  Starting from each task's reference solution, two annotators introduce 1--4 localized edits that preserve executability while causing at least one unit test to fail.

  \item \textbf{\benchmark{}-\textsc{Human-Edited LM}.}
  For each task, we prompt \texttt{gpt-5-mini} to solve the original BigCodeBench problem, without asking it to generate a bug.
  We resample up to $16$ times until obtaining an executable but incorrect program.
  Annotators then edit this draft, for example by changing indices, conditions, or initializations, while keeping it executable and incorrect.
  Tasks for which no executable failing draft is found within the sampling budget are removed.

  \item \textbf{\benchmark{}-\textsc{Qwen-7B}.}
  We prompt \texttt{Qwen2.5-Coder-7B-Instruct} to solve each task and retain one sampled program that executes but fails at least one unit test, resampling up to $16$ times.

  \item \textbf{\benchmark{}-\textsc{gpt-oss-20b}.}
  We apply the same procedure with \texttt{gpt-oss-20b}, yielding errors from a stronger code model.
\end{itemize}

\paragraph{Task alignment and splits.}
To ensure controlled evaluation, we compute the intersection of task IDs across all bug-source variants separately for each split.
We provide a large test split, \texttt{test\_all}, with $517$ examples per source ($2068$ total instances), and a smaller \texttt{test} split with $127$ examples per source ($508$ total instances).
The main experiments use the smaller \texttt{test} split.
We also construct training splits by taking task IDs outside \texttt{test} and \texttt{test\_all} and intersecting them with BigCodeBench-\texttt{train} to avoid leakage.
These training splits are used to form the reference pools for \ourslong{}.

\subsection{Bug Source Analyses}
\label{appendix:benchmark_analyis}%

\begin{figure}[h]
  \centering
  \begin{subfigure}[t]{0.57\textwidth}
    \centering
    \includegraphics[width=\linewidth]{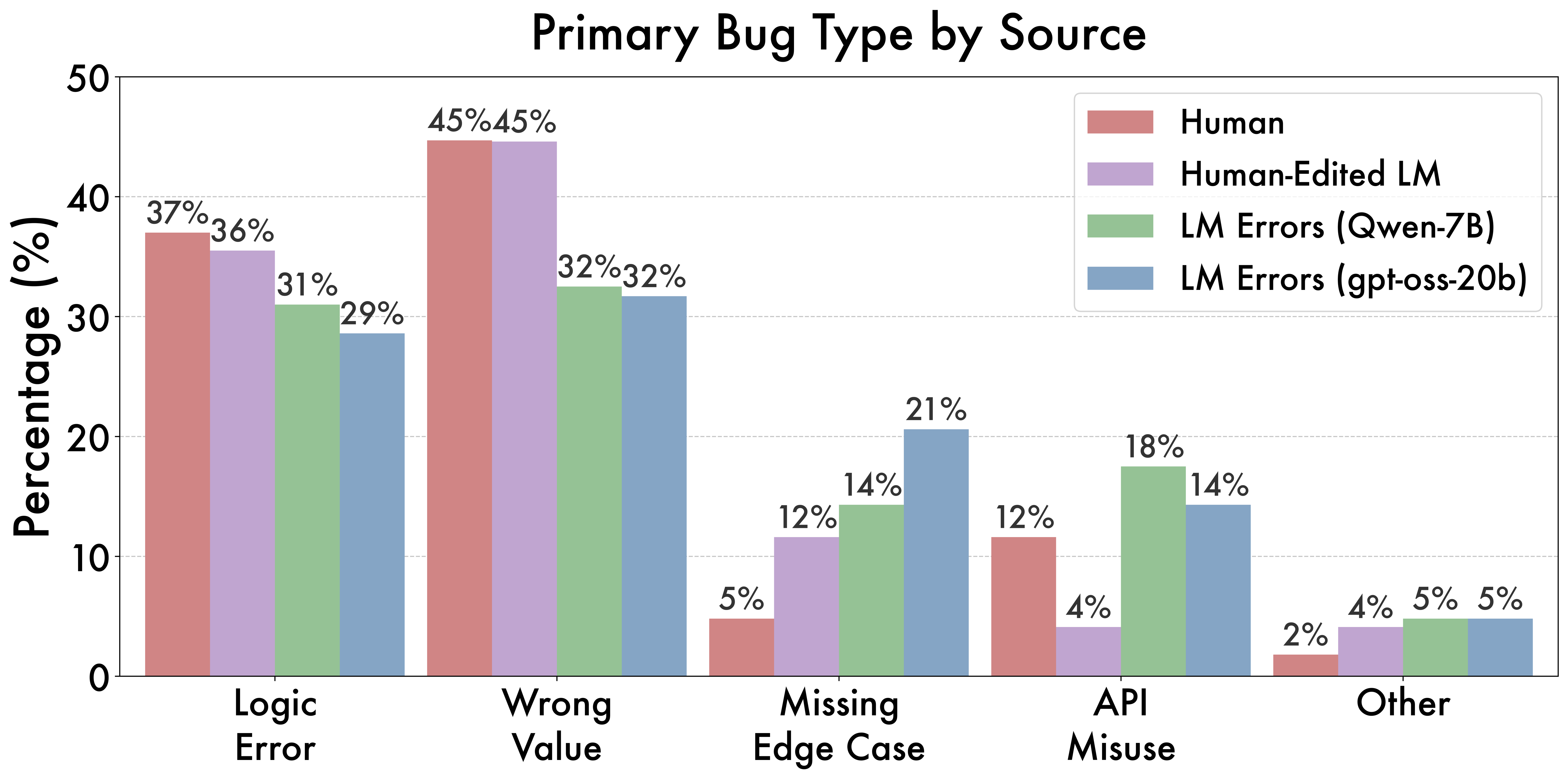}
    \caption{Bug-type profiles by source.}
    
  \end{subfigure}\hfill
  \begin{subfigure}[t]{0.43\textwidth}
    \centering
    \includegraphics[width=\linewidth]{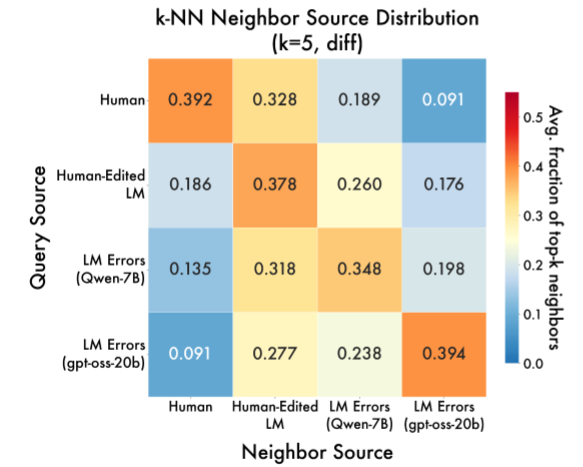}
    \caption{$k$-NN source clustering in embedding space.}
  \end{subfigure}
\caption{
    \textbf{Characterizing bug sources.}
    We label each buggy program with a coarse semantic category using GPT-4o and report the distribution for each \benchmark{} split.
    Bug types vary systematically by source: human-edited bugs skew toward logic errors, gpt-oss-20b bugs toward edge-case and constraint violations, and Qwen-7B bugs toward type and API mistakes.
    We also embed each reference-to-bug diff with \texttt{voyage-code-3} and compute the fraction of its $k$ nearest neighbors, excluding same-task neighbors, that come from each source.
    Diagonal dominance indicates that bugs cluster by source in embedding space.
}
  \label{fig:bug_analysis}
\end{figure}

We characterize bugs along two axes: semantic bug type and embedding-space source structure.
For semantic type, we use five coarse categories:
\textsc{Logic Error}, where the algorithm or reasoning is incorrect;
\textsc{Wrong Value}, where an identifier, literal, constant, return value, or boundary condition is wrong;
\textsc{Missing Edge Case}, where special cases or validations are missing or mishandled;
\textsc{API Misuse}, where a library or framework API is used incorrectly; and
\textsc{Other}, for errors not captured by the previous categories.
We label each buggy program with GPT-4o conditioned on the task specification, reference solution, buggy code, and unit-test traces.
\Cref{fig:bug_analysis} reports the resulting type distribution and an embedding-based source-clustering analysis.

\begin{table}[h]
\centering
\caption{\textbf{Test failure rates by bug source.} We report the mean fraction of unit tests failed by each bug and the percentage of bugs that fail all tests.}
\small
\setlength{\tabcolsep}{6pt}
\begin{tabular}{lcc}
\toprule
\textbf{Source} & \textbf{Mean Fail Rate} & \textbf{100\% Fail} \\
\midrule
Human         & 66.2\% & 31.5\% \\
Human-Edited LM & 54.2\% & 17.3\% \\
LM Errors (Qwen-7B) & 64.5\% & 29.1\% \\
LM Errors (gpt-oss-20b)   & 53.5\% & 20.5\% \\
\bottomrule
\end{tabular}
\label{tab:tabletestcasefail}
\end{table}

\paragraph{Failure-rate analysis.}
\Cref{tab:tabletestcasefail} reports how often bugs from each source fail the unit tests.
\textsc{Human} bugs have the highest mean failure rate and the largest fraction of bugs that fail all tests.
\textsc{Human-Edited LM} bugs are more subtle on average, often passing a larger fraction of the test suite.

\subsection{Evaluation of Frontier Models on \benchmark{}}
\label{appendix:benchmark_abla}

\Cref{tab:frontier} reports fix rates for frontier models across bug sources and repair interfaces.

\begin{table}[h]
\centering
\caption{\textbf{Repair interface comparison.} Fix rate (\%) across \benchmark{} bug sources for \textsc{Repair} (full repair), \textsc{+Tests} (repair with unit-test traces), and \textsc{Diff} (patch output). \textsc{Codegen} is accuracy on the original code-generation tasks.}
\small
\setlength{\tabcolsep}{3pt}
\begin{tabular}{l c ccc ccc ccc ccc}
\toprule
& \multicolumn{1}{c}{\textbf{Codegen}}
& \multicolumn{3}{c}{\textbf{Human}}
& \multicolumn{3}{c}{\textbf{Human-Edited LM}}
& \multicolumn{3}{c}{\textbf{LM (Qwen-7B)}}
& \multicolumn{3}{c}{\textbf{LM (gpt-oss-20b)}} \\
\cmidrule(lr){2-2}\cmidrule(lr){3-5}\cmidrule(lr){6-8}\cmidrule(lr){9-11}\cmidrule(lr){12-14}
\textbf{Model} & \textbf{Score}
& Repair & +Tests & Diff
& Repair & +Tests & Diff
& Repair & +Tests & Diff
& Repair & +Tests & Diff \\
\midrule
\textbf{GPT-5.2}       & 46.5 & 67.7 & \textbf{67.7} & 62.2 & 36.2 & \textbf{53.5} & 32.3 & 19.7 & \textbf{44.9} & 18.1 & 17.3 & \textbf{50.4} & 18.9 \\
\textbf{o4-mini}       & 48.0 & 66.1 & \textbf{70.9} & 55.1 & 37.8 & \textbf{56.7} & 31.5 & 18.9 & \textbf{44.9} &  9.4 & 15.0 & \textbf{49.6} &  9.4 \\
\textbf{Sonnet} & 48.0 & 76.4 & \textbf{81.1} & 72.4 & 39.4 & \textbf{51.2} & 35.4 & 15.7 & \textbf{44.9} & 14.2 & 13.4 & \textbf{44.9} & 14.2 \\
\bottomrule
\end{tabular}
\label{tab:frontier}
\end{table}

\paragraph{Bug-source shift affects repair performance.}
Fix rates vary substantially across bug sources.
Bugs from incorrect model-generated solutions, especially Qwen-7B and gpt-oss-20b, are consistently harder to repair than human-written or human-edited LM bugs.
This supports the need for controlled cross-source evaluation: conclusions from one bug source may not transfer to another.

\subsection{Ablation of Repair Interfaces for \benchmark{}}

We compare four interfaces:
\begin{itemize}[leftmargin=*, nosep]
    \item \textbf{Code generation:} generate a complete solution from the task description.
    \item \textbf{Full repair:} generate a corrected program given the buggy code.
    \item \textbf{Diff repair:} output a patch to modify the buggy code.
    \item \textbf{Repair with test traces:} generate a corrected program given the buggy code and unit-test error traces.
\end{itemize}
Results are shown in \cref{tab:frontier}.
Test traces improve repair performance, while diff repair often underperforms due to brittle formatting and patch application.
For diff repair, we use a custom patch applier with fuzzy context matching.
We use repair with test traces as the default interface in the main experiments because it performs well and matches common test-driven debugging workflows.

\paragraph{Debugging is distinct from code generation.}
Solving a task from scratch and repairing an existing solution succeed on different instances.
\Cref{tab:repair_outcomes_hier} reports the fraction of examples where one interface succeeds and the other fails.
For example, on \benchmark{}-\textsc{Human-Edited LM}, Sonnet solves $41.0\%$ of tasks from scratch that it fails to repair, while repairing $43.9\%$ of tasks that it fails to solve from scratch.
Similar patterns hold across models and bug sources, suggesting that repair requires capabilities, such as fault localization, that are not captured by end-to-end code generation alone.

\begin{table}[h]                                                                                                                                          
  \centering                                                                                                                                                
  \caption{\textbf{Repair and code-generation disagreement rates.}
Entries report the percentage of examples where one interface succeeds and the other fails.}                                                                                                  
  \small                                                                                                                                                    
  \setlength{\tabcolsep}{5pt}                                                                                                                               
  \begin{tabular}{l ccc ccc}
  \toprule                                                                                                                                                  
  & \multicolumn{3}{c}{\textbf{Human}}                                                                                                                      
  & \multicolumn{3}{c}{\textbf{Human-Edited LM}} \\                                                                                                         
  \cmidrule(lr){2-4}\cmidrule(lr){5-7}                                                                                                                      
  \textbf{Outcome}                            
  & \textbf{o4-mini} & \textbf{GPT-5.2} & \textbf{Sonnet}                                                                                                   
  & \textbf{o4-mini} & \textbf{GPT-5.2} & \textbf{Sonnet} \\
  \midrule                                                                                                                                                  
  Fix Fail $\mid$ Code Pass            
  & 16.4 & 16.9 &  8.2                                                                                                                                      
  & 27.9 & 27.1 & 41.0 \\                                                                                                                                   
  Fix Pass $\mid$ Code Fail             
  & 59.1 & 54.4 & 71.2                                                                                                                                      
  & 42.4 & 36.8 & 43.9 \\                                                                                                                                   
  \midrule                                    
  & \multicolumn{3}{c}{\textbf{LM Errors (Qwen-7B)}}                                                                                                                    
  & \multicolumn{3}{c}{\textbf{LM Errors (gpt-oss-20b)}} \\             
  \cmidrule(lr){2-4}\cmidrule(lr){5-7}                                                                                                                      
  & \textbf{o4-mini} & \textbf{GPT-5.2} & \textbf{Sonnet}
  & \textbf{o4-mini} & \textbf{GPT-5.2} & \textbf{Sonnet} \\                                                                                                
  \midrule                                                                                                                                                  
  Fix Fail $\mid$ Code Pass            
  & 42.6 & 39.0 & 41.0                                                                                                                                      
  & 41.0 & 28.8 & 39.3 \\                                                                                                                                   
  Fix Pass $\mid$ Code Fail             
  & 33.3 & 30.9 & 31.8                                                                                                                                      
  & 40.9 & 32.4 & 30.3 \\                                                                                                                                   
  \bottomrule                                 
  \end{tabular}                                                                                                                                             
  \label{tab:repair_outcomes_hier}                          
  \end{table}


\section{Training Details}
\label{sec:appendix-training-details}

\subsection{\ourslong{} and Self-Play Hyperparameters}
\label{appendix:hyperparameters}

We train a \texttt{Qwen2.5-Coder-7B-Instruct} policy with GRPO using the following settings.

\begin{itemize}[leftmargin=*, nosep]
    \item \textbf{Optimization and regularization.}
    Learning rate $1\times 10^{-6}$.
    PPO-style clipping with ratio $0.28$.
    \item \textbf{Batching.}
    Training batch size $64$.
    Validation batch size $256$.
    PPO minibatch size $32$.
    Dynamic batch sizing enabled.
    Maximum PPO token budget of $30{,}000$ tokens per GPU.
    \item \textbf{Sequence lengths.}
    Maximum prompt length $8192$ tokens.
    Maximum response length $2048$ tokens.
    \item \textbf{Rollouts.}
    Asynchronous rollouts.
    Sampling temperature $0.6$ for training and validation.
    Top-$p$ $0.95$ for validation.
    Samples per prompt: $n=4$ for training and $n=1$ for validation.
    \item \textbf{Systems settings.}
    Gradient checkpointing enabled.
    \item \textbf{Training schedule and logging.}
    $10$ epochs total.
    \item \textbf{Parallelism and hardware.}
    One node with $8$ GPUs.
    Tensor parallel size $1$ and sequence parallel size $1$.
\end{itemize}

\paragraph{Self-play loop.}
For each task, we sample $G=4$ candidate bugs.
For each bug, we sample $K=4$ independent repair attempts and estimate its solve rate $\rho$.
The generator receives a band-shaped difficulty reward with $\rho_\ell=0.25$, $\rho_h=0.75$, invalid-bug reward $-1.0$, and extreme-case penalty $\alpha=0.2$.
The fixer receives failing test output as context.
Generator and fixer advantages are normalized separately.

\paragraph{Anchoring.}
For \ourslong{}, the reference pool contains bugs from all four \benchmark{} sources: \textsc{Human}, \textsc{Human-Edited LM}, \textsc{Qwen-7B}, and \textsc{gpt-oss-20b}.
We use \texttt{voyage-code-3} to embed diffs between the reference solution and buggy program.
The generator similarity reward uses $k$-NN scoring with $k=5$, margin scoring with temperature $5.0$, reward weight $\lambda=0.20$, and an exponential-moving-average baseline with decay $\beta=0.99$.
For reference mixing, $20\%$ of fixer-training samples are drawn from the reference pool.

\subsection{Fixer-Only Hyperparameters}

We train a \texttt{Qwen2.5-Coder-7B-Instruct} fixer with GRPO while keeping the generator frozen.
To approximately match compute with Self-Play and \ours{}, we use $n=16$ actor rollouts per training prompt; validation uses $n=1$.

\paragraph{Data and prompting.}
Fixer-Only is trained on a mixture of BigCodeBench tasks and the same reference bug-source training splits available to \ours{}.
During repair, the fixer receives failed unit-test output when available.
At validation time, we evaluate both repair and standard code generation.


\begin{itemize}[leftmargin=*, nosep]
    \item \textbf{Optimization and regularization.}
    Learning rate $1\times 10^{-6}$.
    PPO-style clipping with ratio $0.28$.
    \item \textbf{Batching.}
    Training batch size $64$.
    Validation batch size $256$.
    PPO minibatch size $32$.
    Dynamic batch sizing enabled.
    Maximum PPO token budget of $24{,}000$ tokens per GPU.
    \item \textbf{Sequence lengths.}
    Maximum prompt length $8192$ tokens.
    Maximum response length $2048$ tokens.
    \item \textbf{Rollouts.}
    Asynchronous rollouts.
    Sampling temperature $0.6$ for training and validation.
    Top-$p$ $0.95$ for both the frozen generator and validation sampling.
    Samples per prompt: $n=16$ for training and $n=1$ for validation.
    \item \textbf{Frozen generator configuration.}
    Generator model: Qwen2.5-Coder-7B-Instruct.
    Generation temperature $0.6$ and top-$p$ $0.95$.
    \item \textbf{Training schedule and logging.}
    $10$ epochs total.
    \item \textbf{Parallelism and hardware.}
    One node with $8$ GPUs.
    Tensor parallel size $1$ and sequence parallel size $1$.
\end{itemize}

\paragraph{Compute.}
All runs use one node with $8$ H100 GPUs and take approximately $48$ hours for $120$ training steps.

\section{Prompts}
\label{appendix:prompts}

We use two role-specific prompts: a \emph{bug generator} prompt that injects faults into a correct reference solution, and a \emph{bug fixer} prompt that repairs a buggy program using unit-test feedback when available.
We report the prompts verbatim.
Placeholders such as \texttt{<PROBLEM>}, \texttt{<CORRECT\_CODE>}, \texttt{<BUGGY\_CODE>}, and \texttt{<FAILED\_TEST\_OUTPUT>} are filled at runtime.

\begin{promptbox}{Bug generator prompt}
You are a *bug generator* for Python solutions to competitive programming problems.

You will be given:

1. A problem description.
2. A *correct* reference implementation in Python.

Your task:

- Introduce **one or a few subtle bugs** into the code.
- The resulting code **must still be syntactically valid Python**.
- It should change the behavior so that **at least one unit test fails**.
- Do **not** drastically rewrite the code; keep the overall structure similar.
- Do **not** change the function signature, imports, or I/O format.
- Output **only** the full buggy Python code inside a single ```python``` block.

Problem:
<PROBLEM>

Correct reference implementation:
<CORRECT_CODE>

Now generate the buggy version of this code. Return the entire function with the buggy code inside a ```python``` block:
\end{promptbox}

\begin{promptbox}{Bug fixer prompt}
You are an expert Python debugging assistant.

You will be given:

1. A problem description.
2. A buggy Python implementation that may fail some hidden unit tests.
3. (Optional) Failed unit test output from running the buggy implementation.

Your task:

- Carefully read the code and identify the bug(s).
- Produce a fixed version of the code that makes all unit tests pass.
- Preserve the original function signature, imports, and I/O format.
- Keep the solution reasonably close to the given implementation.
- Output **only** the full corrected Python code inside a single ```python``` block.

Problem:
<PROBLEM>

Buggy implementation:
<BUGGY_CODE>

Failed unit test output (if available):
```text
<FAILED_TEST_OUTPUT>
```
Now fix the bugs in this code. Return the entire function with the fixed code inside a ```python``` block:
\end{promptbox}

\section{Additional Evaluations}
\label{appendix:additional_c}

\subsection{Pass@k Results}
\label{appendix:passk}

We report pass@$k$ under temperature sampling in \cref{tab:passk}, averaging over all \benchmark{} sources.
\ours{} outperforms all baselines at each sampling budget, and the gap increases at larger $k$.

\begin{table}[h]
\centering
\caption{\textbf{Pass@$k$ on \benchmark{}.}
Results are averaged across bug sources.}
\small
\setlength{\tabcolsep}{6pt}
\begin{tabular}{lccc}
\toprule
\textbf{Method} & \textbf{pass@1} & \textbf{pass@5} & \textbf{pass@10} \\
\midrule
Base Model   & 25.8 & 37.3 & 43.8 \\
Fixer-Only   & 27.9 & 40.4 & 46.2 \\
Self-Play    & 29.1 & 41.5 & 48.3 \\
\textbf{\ours{}} (Ours) & \textbf{36.1} & \textbf{50.7} & \textbf{57.2} \\
\bottomrule
\end{tabular}
\label{tab:passk}
\end{table}

\subsection{Qualitative Examples of Generated Bugs}
\label{appendix:ourbugs}

\Cref{fig:example-bugs} compares bugs generated before and after training with \ours{}.

\begin{figure}[h]
\begin{codebox}[left=5pt,right=5pt]{BugSourceBench TaskID 953}
\fontsize{7}{8}\selectfont
Generates random data points to plot bar charts for each in a given list of plot names, then saves them in a specified directory. This function takes a list of plot names, for each generating 10 random data points in [0, 1) to create a bar chart, then saves the bar charts as .png files in the specified directory, creating the directory if it does not exist.\\ 
$>>$ \texttt{task\_func(['First Plot', 'Second Plot'], './another\_folder/')}\\
\texttt{['First\_Plot.png', 'Second\_Plot.png']}\\
\textit{Note that:} This function deduplicates \texttt{mystrings} while maintaining its original order. Random data points for bar charts are generated in the range [0, 1). Each bar chart contains 10 data points.\\
\textit{The function should raise the exception for:} \texttt{FileNotFoundError}: If the provided directory path does not exist and cannot be created.\\
\textit{The function should output with:} list: Names of the files where the plots are saved. Each file corresponds to a title from \texttt{mystrings}.
\end{codebox}
\vspace{-12pt}
\begin{minipage}[t]{0.498\textwidth}
\begin{codebox}{BASE}
\begin{lstlisting}[style=py]
import numpy as np
import matplotlib.pyplot as plt
import os 
def task_func(mystrings, folder_path, seed=None):
    if seed is not None:
        np.random.seed(seed)
    saved_plots = []
    processed_names = set()
    if not os.path.exists(folder_path):
        @@os.mkdir(folder_path)@@
    for name in mystrings:
        if name in processed_names:
            continue
            data = np.random.rand(10)
            plt.bar(range(len(data)), data)
        plt.title(name)
        file_name = name.replace(" ", "_") + ".png"
        plt.savefig(os.path.join(folder_path, file_name))
        saved_plots.append(file_name)
        processed_names.add(name)
    return saved_plots
\end{lstlisting}
\end{codebox}
\end{minipage}\hfill
\begin{minipage}[t]{0.498\textwidth}
\begin{codebox}{ASP}
\begin{lstlisting}[style=py]
import numpy as np
import matplotlib.pyplot as plt
import os
def task_func(mystrings, folder_path, seed=None):
    if seed is not None:
        np.random.seed(seed)
        
    saved_plots = []
    if not os.path.exists(folder_path):
        @@raise Exception("Directory could not be created.")@@ 
    
    for name in mystrings:
        @@data = np.random.rand(10)@@
        plt.bar(range(len(data)), data)
        plt.title(name)
        file_name = name.replace(" ", "_") + ".png"
        plt.savefig(os.path.join(folder_path, file_name))
        saved_plots.append(file_name)
        
    @@return ["error"]@@
    
\end{lstlisting}
\end{codebox}
\end{minipage}
\vspace{1pt}

\caption{\benchmark{} example comparing bugs generated by the base model and by the \ours{}-trained generator.
The base generator produces a simple synthetic error, while the \ours{}-trained generator introduces more semantic failures, including missing de-duplication, improper error handling, and an incorrect return value.}
\label{fig:example-bugs}
\end{figure}

\subsection{Similarity of Generated Bugs}
\label{sec:target_like_bugs}

\paragraph{Mean similarity over training.}
For each generator checkpoint, we sample synthetic bugs on held-out \benchmark{} tasks.
We embed each reference-to-bug diff and compute its mean top-$k$ cosine similarity to a target bug pool, excluding same-task matches when available.
Averaging this score across generated bugs gives a similarity trajectory that measures how the generator distribution moves relative to the target source over training.

\paragraph{Fix rate by generation-similarity quantile.}
To test whether downstream gains are explained only by producing more target-like bugs, we bin tasks by the generator's similarity to the target pool using shared quantile boundaries across methods.
Within each bin, we evaluate fixer performance on the corresponding held-out \benchmark{} instances.
This matched-similarity diagnostic compares \ours{} and Self-Play at comparable levels of generator-target similarity, allowing us to test whether \ours{} improves robustness beyond shifting the generator toward higher-similarity bugs.

\begin{figure}[h]
\centering
    \begin{subfigure}[t]{0.49\textwidth}
        \centering
        \includegraphics[width=\linewidth]{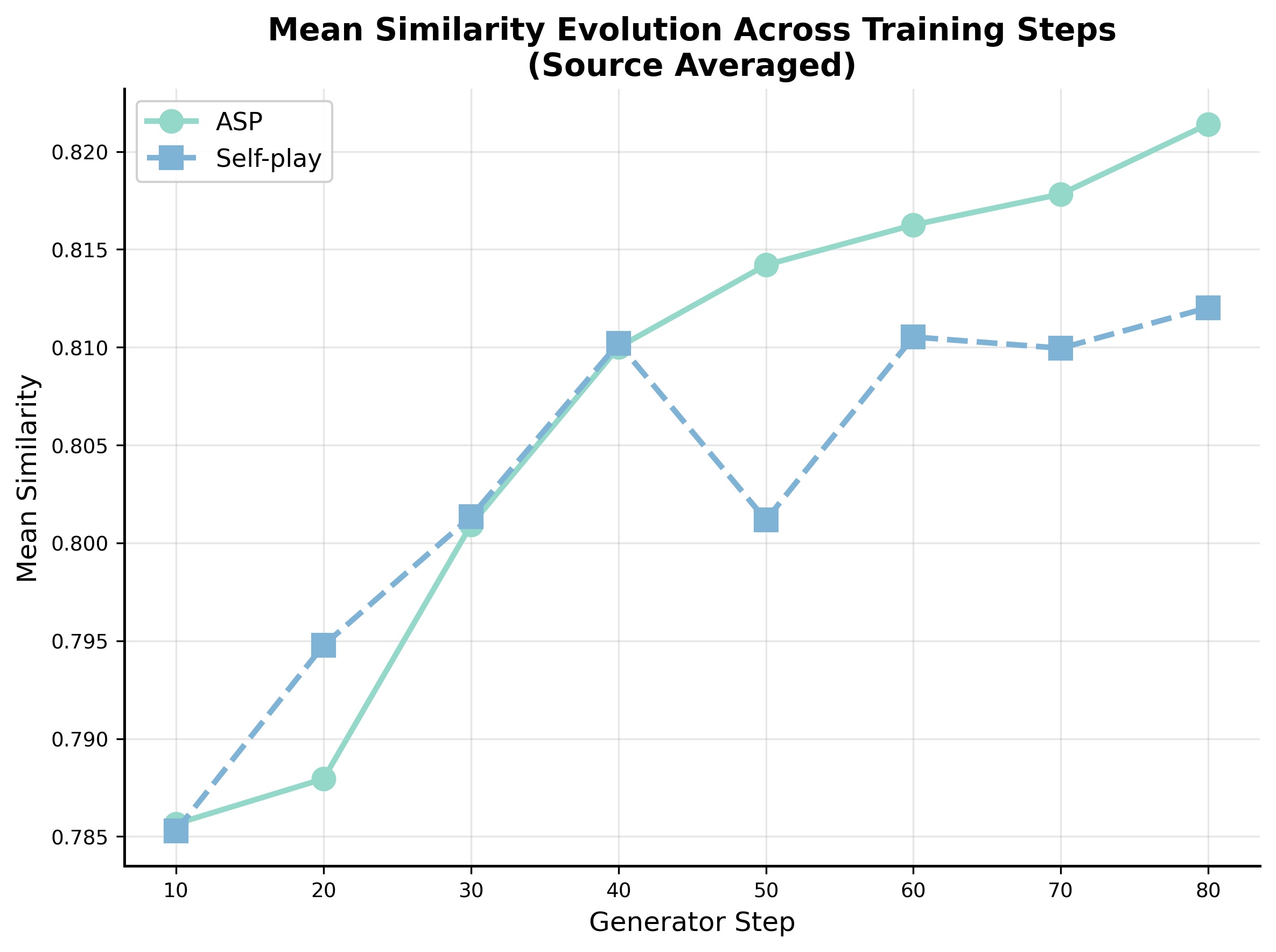}
  \caption{\textbf{Similarity-guided shaping increases target-likeness of generated bugs.}}
  \label{fig:code_similarity_plot}    
    \end{subfigure}\hfill
    \begin{subfigure}[t]{0.49\textwidth}
        \centering
        \includegraphics[width=\linewidth]{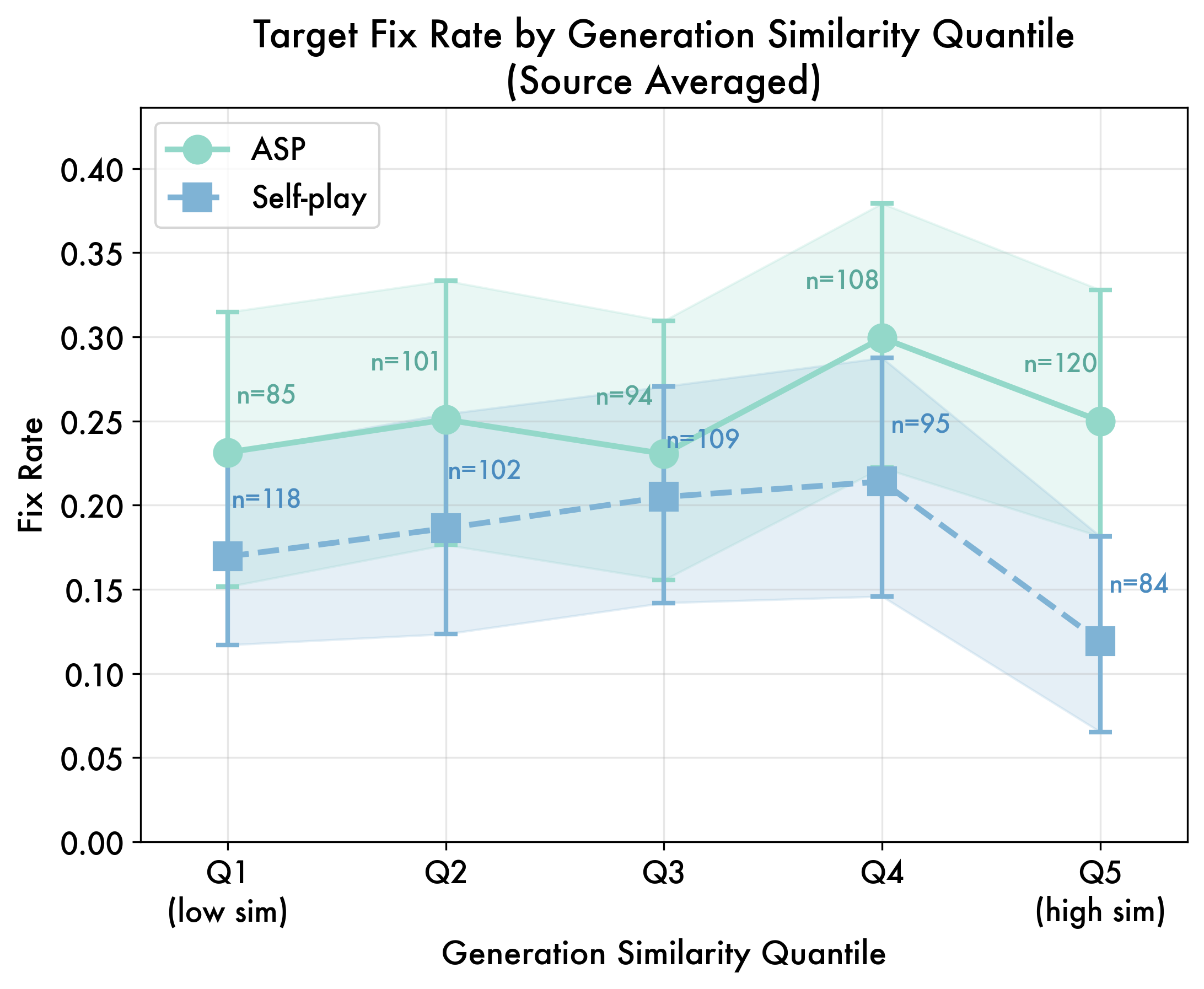}
        \caption{\textbf{Performance on test tasks by generated similarity.}}
    \end{subfigure}
    \caption{
We sample $n=3$ bugs from the generator for each held-out \benchmark{} task and compute the $k$-NN embedding similarity of generated bugs to target-source bugs.
Similarity-guided shaping in \ours{} increases target similarity over training relative to standard self-play (left).
We then bin tasks by generated-bug similarity, sample $k=3$ repair attempts per task, and report fix rate with 95\% confidence intervals (right).
}
\label{fig:similarity_and_buckets}
\end{figure}

\subsection{Evaluation by Semantic Bug Type}
\label{appendix:bugtype_eval}

We group bugs by semantic type and report fix rates in \cref{tab:bugtype_eval}.
\ours{} improves most categories, suggesting that its gains are not limited to a single error pattern.

\begin{table}[h]
\centering
\caption{\textbf{Fix rate (\%) by semantic bug type.} Fix rates are reported in percent.}
\small
\setlength{\tabcolsep}{5pt}
\begin{tabular}{lccccc}
\toprule
\textbf{Model} & \textbf{Logic} & \textbf{Wrong Val.} & \textbf{Edge Case} & \textbf{API Misuse} & \textbf{Other} \\
\midrule
Qwen-7B (base)  & 14 & 37 &  5 & 20 & 33 \\
Fixer-Only       & 12 & 37 &  5 & 14 & 28 \\
Self-Play        & 16 & 37 &  9 & 18 & 44 \\
\textbf{\ours{}} (Ours) & \textbf{21} & 36 & \textbf{12} & \textbf{29} & 38 \\
\bottomrule
\end{tabular}
\label{tab:bugtype_eval}
\end{table}

\subsection{Evaluation as a Test-Time Fixer}
\label{appendix:testtime_fixer}

We evaluate whether \ours{}-trained fixers can improve larger code-generation models at test time.
We use DeepSeek-Coder-33B and Qwen2.5-Coder-32B as initial coders, generate candidate solutions, and then apply 7B fixer models for $r$ repair rounds.
\Cref{tab:testtime_fixer} reports pass rates with $k$ samples.
\ours{} generally outperforms the base, Fixer-Only, and Self-Play fixers, and can improve over self-repair by the larger coder.

\begin{table}[h]                                               \centering                                                     \caption{\textbf{Test-time fixer results.} Pass rate (\%) using large coders with fixer models over $r$ repair rounds and $k$ samples. 
  The \emph{No repair} row is the coder alone ($r{=}0$); all other rows apply the listed fixer.}                          \small                                                                                                                                              
  \setlength{\tabcolsep}{5pt}                                                                                                                               
  \begin{tabular}{lcccc}                                                                                                                                    
  \toprule                                                                                                                                                  
   & \multicolumn{2}{c}{$r{=}1$} & \multicolumn{2}{c}{$r{=}2$} \\
  \cmidrule(lr){2-3}\cmidrule(lr){4-5}                                                                                                                      
  \textbf{Fixer} & $k{=}1$ & $k{=}2$ & $k{=}1$ & $k{=}2$ \\                                                                                                 
  \midrule
  \multicolumn{5}{l}{Coder: DeepSeek-33B \hfill \textit{No repair: $k{=}1$: 29.5 \quad $k{=}2$: 37.0}} \\                                          
  \midrule                                                                                                                                                  
  Qwen-7B (base)          & 35.4 & 40.9 & 36.2 & 47.2 \\                                                                                                    
  Fixer-Only               & 34.6 & 41.7 & 35.4 & 41.7 \\                                                                                                   
  Self-Repair (Qwen-32B)   & 34.6 & 42.5 & 37.8 & 49.6 \\                                                                                                   
  Self-Play                & 35.4 & 42.5 & 37.8 & 46.5 \\   
  \textbf{\ours{}} (Ours)  & \textbf{36.2} & \textbf{45.7} & \textbf{43.3} & \textbf{53.5} \\                                                               
  \midrule                                                                                                                                                  
  \multicolumn{5}{l}{Coder: Qwen-32B \hfill \textit{No repair: $k{=}1$: 30.3 \quad $k{=}2$: 37.0}} \\                                              
  \midrule                                                                                                                                                  
  Qwen-7B (base)               & 33.1 & 40.9 & 36.2 & 44.1 \\
  Fixer-Only                    & 33.9 & 40.2 & 37.0 & 43.3 \\                                                                                              
  Self-Repair (DeepSeek-33B)    & 35.4 & 40.9 & \textbf{44.9} & \textbf{57.5} \\
  Self-Play                     & 33.9 & 43.3 & 37.8 & 48.0 \\
  \textbf{\ours{}} (Ours)       & \textbf{38.6} & \textbf{44.1} & \textbf{44.9} & 54.3 \\                                                          
  \bottomrule                                                                                                                                               
  \end{tabular}                                             
  \label{tab:testtime_fixer}                                                                                                                                
  \end{table}                                               

\FloatBarrier
\section{Additional Experiments}
\label{appendix:codegengrpo_c2}

\subsection{Additional Baselines}
\label{appendix:more_baselines}

We compare against additional baselines that test alternative ways of improving repair.

\paragraph{RL for code generation.}
To test whether improvements come from better general code-generation ability rather than repair-specific training, we train the base model to generate solutions from task specifications using binary unit-test rewards.
We use the same task splits and comparable training settings as in the repair experiments.

\paragraph{Self-play with constrained bug generation.}
As an alternative to open-ended bug generation, we evaluate a constrained generator based on AST mutations.
We identify valid mutation sites in the reference implementation and ask the generator to select mutations that preserve syntactic validity while producing failing programs.
We then train with the same self-play procedure.

\paragraph{Supervised fine-tuning for repair.}
We train a fixer with supervised fine-tuning on \benchmark{} training instances from all four sources, using a masked cross-entropy loss on the reference correct solutions.

\paragraph{Alternate embeddings for \ours{}.}
To test sensitivity to the embedding model, we run \ours{} with \texttt{CodeBERT} embeddings instead of \texttt{voyage-code-3}.

\begin{table}[h]
\centering
\caption{\textbf{Additional baselines.}
For \benchmark{} splits, we report repair fix rate (\%).
For \textsc{Codegen}, we report pass rate on the original code-generation task.}
\small
\setlength{\tabcolsep}{4pt}
\begin{tabular}{lccccc}
\toprule
\textbf{Method} & \textbf{Codegen} & \textbf{Human} & \textbf{Human-Ed.} & \textbf{LM (Qwen-7B)} & \textbf{LM (gpt-oss-20b)} \\
\midrule
Base Model                  & 21.7 & 58.3 & 30.8 &  7.1 &  7.0 \\
Codegen RL & \textbf{31.8} & 55.8 & 30.8 & 2.9 & 6.3 \\
AST-Constrained Self-Play & -- & 59.8 & 26.8 & 3.9 & 3.1 \\
Fixer SFT & -- & 46.5 & 18.1 & 10.2 & 11.0 \\
\ours{} (\texttt{CodeBERT}) & -- & \textbf{63.3} & \textbf{33.9} & \textbf{23.6} & \textbf{24.6} \\
\bottomrule
\end{tabular}
\label{tab:codegengrpo}
\end{table}

Results in \cref{tab:codegengrpo} show that RL post-training for code generation improves code-generation pass rate but does not improve repair, and often degrades it.
Fixer SFT improves the LM-generated splits slightly but substantially degrades Human and Human-Edited LM performance.
The \texttt{CodeBERT} variant of \ours{} remains strong, suggesting that the method is not tied to a single embedding model.

\subsection{Additional Ablations}
\label{appendix:more_ablations}

\paragraph{Embedding and similarity ablations.}
We ablate the $k$-NN pooling parameter and the embedding model used for the similarity reward in (\cref{tab:knn_k}).
\ours{} is stable across moderate values of $k$ and performs similarly with \texttt{voyage-code-3} and \texttt{CodeBERT} embeddings.

\begin{table}[h]
\centering
\caption{\textbf{Embedding and $k$-NN ablations.}
We vary the $k$-NN pooling parameter using \texttt{voyage-code-3} embeddings and compare against \texttt{CodeBERT}.}
\small
\begin{tabular}{lc}
\toprule
\textbf{$k$} & \textbf{Avg Fix Rate (\%)} \\
\midrule
1              & 34.7 \\
5 (default)    & 36.1 \\
10             & 35.8 \\
20             & 35.0 \\
\bottomrule
\end{tabular}
\quad
\begin{tabular}{lc}
\toprule
\textbf{Embedding} & \textbf{Avg Fix Rate (\%)} \\
\midrule
voyage-code-3 (default) & 36.1 \\
CodeBERT                & 36.4 \\
\bottomrule
\end{tabular}
\label{tab:knn_k}

\end{table}

\paragraph{Updating the generator on mixed episodes.}
Reference mixing replaces a generated bug with a reference bug for fixer training.
We do not update the generator on these mixed episodes because the bug was not sampled from the generator; applying a generator update would be off-policy or require additional fixer rollouts.
\Cref{tab:gen_on_mixed} shows that updating the generator on mixed episodes reduces average fix rate from $36.1\%$ to $33.6\%$, supporting our choice to update only the fixer on reference-mixed episodes.

\begin{table}[h]
\centering
\caption{\textbf{Updating the generator on reference-mixed episodes.}
Fix rates are reported on \benchmark{} splits.}
\small
\setlength{\tabcolsep}{4pt}
\resizebox{0.6\columnwidth}{!}{%
\begin{tabular}{lccccc}
\toprule
\textbf{Method} & \textbf{Human} & \textbf{Human-Ed.} & \textbf{LM (Qwen-7B)} & \textbf{LM (gpt-oss)} & \textbf{Avg} \\
\midrule
Base Model & 58.3 & 30.8 &  7.1 &  7.0 & 25.8 \\
Fixer-Only & 58.3 & 30.7 & 10.2 & 12.6 & 27.9 \\
Self-Play  & 63.7 & 31.5 &  8.7 & 12.6 & 29.1 \\
\textbf{\ours{}} (Ours) & \textbf{65.0} & \textbf{37.0} & \textbf{21.3} & \textbf{21.3} & \textbf{36.1} \\
\ours{} + gen on mixed & 64.1 & 35.8 & 16.1 & 18.3 & 33.6 \\
\bottomrule
\end{tabular}}
\label{tab:gen_on_mixed}
\end{table}

\paragraph{Decoupled generator and fixer.}
We test whether drift and \ours{}'s gains depend on sharing weights between generator and fixer.
\Cref{tab:decoupled} compares shared and decoupled variants.
Drift is not removed by decoupling, and \ours{} remains strong with separate generator and fixer weights.
This suggests that drift arises from the unit-test-only reward signal rather than from weight sharing alone.

\begin{table}[h]
\centering
\caption{\textbf{Shared vs.\ decoupled generator/fixer weights.}
We compare standard Self-Play and \ours{} with shared or separate generator and fixer models.}
\small
\setlength{\tabcolsep}{5pt}
\begin{tabular}{llccccc}
\toprule
\textbf{Method} & \textbf{Weights} & \textbf{Avg} & \textbf{Human} & \textbf{Human-Edited LM} & \textbf{LM (Qwen-7B)} & \textbf{LM (gpt-oss-20b)} \\
\midrule
Self-Play & shared    & 20.9 & 31.7 & 24.4 & 12.6 & 15.0 \\
Self-Play & decoupled & 19.6 & 28.5 & 22.5 & 12.0 & 15.5 \\
\ours{}   & shared    & 35.5 & 64.2 & 37.0 & 18.9 & 22.0 \\
\ours{}   & decoupled & 34.8 & 61.5 & 36.0 & 18.0 & 23.5 \\
\bottomrule
\end{tabular}
\label{tab:decoupled}
\end{table}

\paragraph{Alternate base model.}
We replicate the main experiments using DeepSeek-Coder-6.7B-Instruct as the base model.
As shown in \cref{tab:alt_base}, Self-Play yields limited gains and slightly degrades human-originated bugs, while \ours{} improves all sources and gains $+8.9$ pp over Self-Play on average.

\begin{table}[h]
\centering
\caption{\textbf{Alternate base model.}
Results using DeepSeek-Coder-6.7B-Instruct as the base model.}
\small
\setlength{\tabcolsep}{5pt}
\begin{tabular}{lccccc}
\toprule
\textbf{Method} & \textbf{Human} & \textbf{Human-Edited LM} & \textbf{LM (Qwen-7B)} & \textbf{LM (gpt-oss-20b)} & \textbf{Avg} \\
\midrule
Base Model   & 51.6 & 22.8 &  3.7 &  2.6 & 20.2 \\
Fixer-Only   & 51.6 & 22.7 &  6.8 &  8.2 & 22.3 \\
Self-Play    & 50.8 & 22.0 &  5.3 &  8.2 & 21.6 \\
\textbf{\ours{}} (Ours) & \textbf{58.3} & \textbf{29.0} & \textbf{17.9} & \textbf{16.9} & \textbf{30.5} \\
\bottomrule
\end{tabular}
\label{tab:alt_base}
\end{table}

\paragraph{External benchmarks and task distributions.}
We also evaluate on DebugBench \citep{tian2024debugbenchevaluatingdebuggingcapability}, a debugging benchmark derived from LeetCode-style problems.
We use the $1{,}340$ Python buggy programs and the same fixer prompt format as in \benchmark{}.
\Cref{tab:debugbench} shows that \ours{} transfers to this external distribution and outperforms standard Self-Play.

\FloatBarrier
\begin{table}[!htbp]
\centering
\small
\captionsetup{font=small}
\begin{subtable}[t]{0.62\textwidth}
\centering
\caption{\textbf{Fix rates on DebugBench.}
Models are trained on BigCodeBench-based tasks and evaluated on DebugBench Python bugs.}
\setlength{\tabcolsep}{6pt}
\begin{tabular}{lc}
\toprule
\textbf{Method} & \textbf{DebugBench}  \\
\midrule
Base Model & 55.0 \\
Codegen RL & 54.3 \\
AST-Constrained Self-Play & 56.6 \\
Fixer SFT & 56.0 \\
Fixer-Only & 54.6 \\
Self-Play & 58.7 \\
\ours{} (\texttt{CodeBERT}) & 59.0 \\
\textbf{\ours{}} (Ours) & \textbf{60.3} \\
\bottomrule
\end{tabular}
\label{tab:debugbench}
\end{subtable}
\hfill
\begin{subtable}[t]{0.34\textwidth}
\centering
\caption{\textbf{Held-out DebugBench.}
Models are trained and evaluated on DebugBench splits.}
\setlength{\tabcolsep}{5pt}
\begin{tabular}{lc}
\toprule
\textbf{Method} & \textbf{Fix} \\
\midrule
Qwen-7B (base) & 58.0 \\
Self-Play       & 58.7 \\
\textbf{\ours{}} (Ours) & \textbf{66.3} \\
\bottomrule
\end{tabular}
\label{tab:debugbench_train}
\end{subtable}

\vspace{2pt}
\caption{\textbf{DebugBench transfer and in-domain training.}
\ours{} improves over standard Self-Play both when transferred from BigCodeBench-based training to DebugBench and when trained directly on DebugBench.}
\label{tab:debugbench_results}
\end{table}